\documentclass[final,1p,times]{elsarticle}

\usepackage{graphicx}
\usepackage{amssymb}
\usepackage{newtxtext}
\usepackage{newtxmath}
\usepackage{mathtools}
\usepackage{amsmath,bm}
\usepackage{tabularx,booktabs}
\usepackage{caption}
\usepackage{rotating}
\usepackage{pdflscape}
\usepackage{xcolor}
\usepackage{overpic}
\usepackage{hyperref}
\usepackage[numbers]{natbib}
\usepackage{booktabs}
\usepackage{siunitx}
\usepackage{tabularx, booktabs, array}
\usepackage{pdflscape}
\usepackage{soul}
\usepackage{lineno}

\newcommand{\f}{\frac}
\newcommand{\p}{\partial}
\newcommand{\bs}[1]{\boldsymbol{#1}}
\newcommand{\be}{\begin{equation}}
\newcommand{\ee}{\end{equation}}
\newcommand{\bse}{\begin{subequations}}
\newcommand{\ese}{\end{subequations}}
\newcommand{\bq}{\bs{q}}
\newcommand{\R}{Re}
\newcommand{\xo}{x_\text{o}}
\newcommand{\td}{\mathrm{d}}
\newcommand{\ti}{\textnormal{i}}

\title{Physically Consistent Outflow Boundary Conditions for\\ Global Stability Analysis of Bluff Body Wakes}

\author{Guangyao Cui}
\author{Amit Sigawi}
\author{Michael Karp\corref{cor1}}

\address{The Stephen B. Klein Faculty of Aerospace Engineering, Technion Israel Institute of Technology, Haifa 3200003, Israel}
\cortext[cor1]{Corresponding author: mkarp@technion.ac.il}

\journal{Journal of Computational Physics}

\begin{document}
\makeatletter
\def\ps@pprintTitle{%
 \let\@oddhead\@empty
 \let\@evenhead\@empty
 \let\@oddfoot\@empty
 \let\@evenfoot\@empty
}
\makeatother

\begin{frontmatter}

\begin{abstract}
Global linear stability analysis of bluff body wake flows is performed using the matrix-forming method based on finite-difference discretization. Particular emphasis is placed on the influence of outflow boundary conditions, with the aim of minimizing the required computational domain size without degrading accuracy or inducing spurious oscillations near the outlet. This study focuses on incompressible wakes behind bluff bodies such as cylinders and airfoils at high angle of attack, especially in regimes where global modes exhibit downstream spatial amplification. It is shown that below the critical Reynolds number -- where the global mode remains linearly stable -- significant spatial growth can persist far downstream, even when the wake is nearly absent. This behavior underscores the importance of imposing a physical boundary condition at the outlet. Several commonly used outflow boundary conditions are evaluated, including Dirichlet, Neumann, extrapolation, stress-free, sponge layer, and the Robin condition that incorporates predictions from local linear stability analysis at the outlet. The results demonstrate that, for different $\R$ cases, the Robin condition enables robust convergence of global modes within substantially truncated domains, thereby improving the efficiency of global stability analysis. These findings highlight the broader applicability of the matrix-forming approach for complex stability analyses, including Floquet analysis of time-periodic flows and extensions to compressible configurations.

\end{abstract}

\end{frontmatter}

\section{Introduction}
\label{sec:intro}

The global linear instability of bluff-body wakes is fundamental to understanding the onset of periodic flow behavior, as it determines the unsteady dynamics of wakes behind cylinders, airfoils, and other bluff bodies, with significant consequences for transition modeling and flow-control design. The seminal work of \citet{vonKarman1911} first showed the formation of a vortex street in the wake of a cylinder, laying the foundation for understanding wake periodicity and its hydrodynamic stability. \citet{barkley1996three}, \citet{barkley2006linear}, and \citet{williamson1989oblique} provided critical insights into the onset of two-dimensional and three-dimensional wake instabilities, identifying the critical Reynolds numbers via linear stability analysis. 

One of the biggest challenges in analyzing the global instabilities of wake flows is the implementation of the outflow boundary condition (OBC), where the complex vortex dynamics is unknown; however, providing an OBC is mathematically necessary. Significant efforts have been made to develop effective OBCs, as reviewed by \citet{sani1994resume} who explained that an ideal OBC should allow the flow to exit the domain without affecting the flow close to it or introducing numerical reflections upstream, and the solution should be independent of the outlet where the computational domain is truncated. \citet{sani1994resume} highlighted the trade-offs among different methods, and noted that no single OBC universally outperforms others, and the choice of OBC and the ease of implementation depend on the problem itself and the discretization scheme. The traction-free OBC \cite{liu2009open, guermond2005error} is often applied in the finite element method to prevent reflections, while the finite difference and finite volume methods often choose convective boundary conditions \cite{orlanski1976simple, ruith2004development} to avoid numerical artifacts. Subsequent studies have proposed additional OBCs, including non-reflecting boundary conditions \cite{jin1993nonreflecting}, no-boundary conditions \cite{papanastasiou1992new}, etc. \citet{dong2014robust} introduced a robust OBC designed for incompressible flow simulations on severely truncated domains that allowed strong vortices to exit without causing numerical instability. The condition is formulated within a rotational velocity correction scheme, incorporating an energy-stable approach that allows kinetic energy influx through the outflow boundary while preventing uncontrolled energy growth.		 

The studies discussed above have contributed to enhancing the efficiency of wake flow simulations and reducing the computational domain required to solve the nonlinear Navier-Stokes equations (NSE). However, applying OBCs in global stability analyses---where the objective is to compute eigenvalues of the linearized Navier-Stokes equations (LNSE)---introduces additional complexities and challenges. The implementation of those OBCs in the linear analysis is non-trivial and often requires adding additional algorithmic complexity, and can possibly introduce spurious modes. Therefore, classical Dirichlet and Neumann OBCs are still widely applied on the outflow boundary in the global stability analysis, with the need to compensate by a very large computational domain for the eigenfunction, consisting of vortices generated upstream, to be sufficiently dissipated before reaching the outflow boundary. 

\citet{barkley1996three} conducted a linear stability analysis of the unsteady cylinder wake flow for Reynolds numbers between 140 and 300, and identified two distinct modes of secondary instability: Mode A, which emerges at a critical Reynolds number of $\R \approx 188.5$, and Mode B, which becomes unstable at $\R \approx 259$ with a shorter spanwise wavelength. In their work, \citet{barkley1996three} applied the Neumann BC for the velocity eigenmode and the Dirichlet BC for the pressure mode in the outlet, and the same OBC was applied in subsequent studies such as \citet{barkley2006linear}, \citet{giannetti2007structural}, and \citet{he2017linear}. The stress-free OBC, which combines the velocity and pressure term, has also been applied in many studies with a slight variation in the definition of the stress tensors (e.g.,~\citet{noack2003hierarchy, mittal2010stability, sen2011flow, sipp2007global}). \citet{canuto2015two} investigated the instabilities of cylinder wakes in compressible flows and added a sponge layer near the outlet to overcome potential issues that may arise from the outlet. However, those OBCs do not necessarily capture the true physics in the outlet, where the actual mode may still be evolving spatially, even very far from the bluff body. How non-physical outflow boundary conditions influence global stability analysis remains insufficiently understood, raising the question of how to impose a more physically consistent outlet condition within the linear eigenvalue framework. A key open question is whether information from the local instability characteristics of the flow at the outlet can be incorporated into the outflow boundary condition used in the global analysis.

In the global stability analysis of a flat plate boundary layer, \citet{ehrenstein2005two} and \citet{alizard2007spatially} proposed the Robin-type convective boundary condition that accounts for Tollmien-Schlichting waves at the inlet and outlet. The streamwise wavenumber can be obtained from the local stability analysis by solving the Orr-Sommerfeld equation. A Taylor expansion was conducted at the neutral point to associate the local wavenumber and global frequency (noting that only the real part in the group velocity was taken into account), and thereby the Robin boundary condition was obtained. A similar OBC that incorporates the `local' wavenumber at the outlet was also applied in \citet{fasel1990numerical}. These studies introduced a novel approach to developing more physical OBCs for global analysis by leveraging insights from local analysis at the outlet. However, most global stability studies still tend to choose the classical Neumann or Dirichlet OBC due to its simple implementation, with the assumption that a large enough domain could compensate for disturbances that may originate from the slightly non-physical boundary conditions (e.g.,~\citet{he2017linear}).

However, the impact of different OBCs on the required domain size in the global stability analysis and how much these choices are affected by parameters such as the Reynolds number remain unclear (e.g.,~\citet{alizard2007spatially, barkley2006linear, mittal2010stability, he2017linear}). Furthermore, most of the studies above used the time-stepping method to solve the global stability problem; however, these OBCs might not be directly applicable to matrix-forming methods \cite{juniper2014modal}. In matrix-forming methods, which usually rely on finite-difference discretizations, the OBC must be incorporated into the discretized operators, which may pose unique implementation challenges. Appropriate boundary conditions must be compatible with the matrix structure, avoid introducing ill-conditioning, and accurately represent the physical outflow without generating spurious modes. Implementing non-reflecting or energy-based OBCs in this framework is nontrivial, as these conditions often involve complex operators that are difficult to discretize consistently within a finite-difference scheme. Further discussion of time stepping and matrix-forming methods can be found in \citet{juniper2014modal}. 

In this study, we analyze the impact of OBCs on the global stability analysis of the two-dimensional wakes of the cylinder and airfoil at a high angle of attack using the matrix-forming method, discretized by the finite-difference scheme. The rest of the paper is organized as follows: In \S\ref{sec:global-stability-formulation}, we first explain the formulation of global stability analysis and develop the global solver for the eigenvalue problem, followed by the base flow solution; the commonly applied boundary conditions are then summarized. In \S\ref{sec:results}, the spatial convergence of the spectrum and eigenfunctions obtained by different OBCs, including Neumann, Dirichlet, extrapolation, stress-free, sponge layer, and two Robin conditions, are evaluated for both cylinder and airfoil wakes. We show that compared to the velocity component, the pressure component of the eigenmode is more difficult to converge and, therefore, requires more effort in the global stability analysis. The impact of the Reynolds number on the behavior of different OBCs and how it is related to the spatial and temporal growth/decay are then explained. Finally, the conclusions are given in \S\ref{sec:conclusion}. This study demonstrates that the Robin boundary condition, by incorporating local linear stability theory, provides a physically consistent outflow treatment that enables the computation of converged global eigenmodes within significantly reduced computational domains. Our results further highlight the potential applicability of the matrix-forming approach to complex stability analyses.

\section{Global Stability Analysis Formulation and Boundary Conditions}
\label{sec:global-stability-formulation}

\subsection{Matrix-Forming Formulation of Global Stability Analysis}

Linear stability analysis is based on the decomposition of the flow variables, $\bs{q}=(u,v,w,p)^\textnormal{T}$, into the base flow, $\overline{\bs{q}}$, and the small perturbation, $\tilde{\bs{q}}$,
\be
	\bs{q}(\bs{x}, t) = \overline{\bs{q}}(\bs{x}) + \varepsilon \tilde{\bs{q}}(\bs{x}, t),\quad \varepsilon\ll1,
\ee
where $\bs{x} = (x,y,z)$ and $t$ correspond to the spatial coordinates (streamwise, transverse, and spanwise) and time, respectively. The global stability analysis can therefore be formulated by substituting this ansatz into the incompressible Navier-Stokes equations (NSE) and neglecting the higher-order perturbation terms, yielding the linearized Navier-Stokes equations (LNSE),

\bse\label{eq:LNSE}
\be\label{eq:LNSE_mom}
	\f{\p \tilde{\bs{u}}}{\p t}+\tilde{\bs{u}} \cdot \nabla \overline{\bs{U}}+\overline{\bs{U}} \cdot \nabla \tilde{\bs{u}} = -\nabla \tilde{p}+\f{1}{\R} \nabla^{2} \tilde{\bs{u}},
\ee
\be\label{eq:LNSE_cont}
	\nabla \cdot \tilde{\bs{u}} = 0,
\ee
\ese
where the base flow is $\overline{\bs{q}}=(\overline{\bs{U}},\overline{P})^\textnormal{T}$ and the perturbation is $\tilde{\bs{q}}=(\tilde{\bs{u}},\tilde{p})^\textnormal{T}$. When the base flow is two-dimensional, i.e., $\overline{\bs{U}}=(\overline{U}(x,y),\overline{V}(x,y),0)^\textnormal{T}$, the perturbation takes the form
\be
	\tilde{\bs{q}}(x,y,z,t) = \hat{\bs{q}}(x,y) e^{\ti(\beta z - \omega t )},
\ee
where $\beta$ is the spanwise wavenumber ($\beta = 0$ for the two-dimensional analysis), and $\omega$ represents the temporal frequency. Substituting into the LNSE equation~\eqref{eq:LNSE} leads to a BiGlobal temporal eigenvalue problem,

\be\label{eq:matrix-form}
	\left[\begin{array}{cccc}
		\mathcal{L} + \overline{U}_x & \overline{U}_y	& 0                   & \mathcal{D}_x \\
		\overline{V}_x	& \mathcal{L} + \overline{V}_y	& 0 		 	      & \mathcal{D}_y \\
		0				& 0								& \mathcal{L}         & \ti\beta\mathcal{I} \\
		\mathcal{D}_x	& \mathcal{D}_y 				& \ti\beta\mathcal{I} & 0
	\end{array}\right]\hat{\bs{q}}=\ti\omega\left[\begin{array}{cccc}
		\mathcal{I} & 0 & 0 & 0 \\
		0 & \mathcal{I} & 0 & 0 \\
		0 & 0 & \mathcal{I} & 0 \\
		0 & 0 & 0 & 0
	\end{array}\right]\hat{\bs{q}},
\ee
where $\mathcal{L} = \overline{U}\mathcal{D}_x + \overline{V}\mathcal{D}_y - {\R}^{-1}(\mathcal{D}_{xx}+\mathcal{D}_{yy} - \beta^2 \mathcal{I})$, and the spatial derivative operators are indicated by subscripts. Comprehensive discussions on modal stability theory were reviewed in \citet{SchmidHenningson2001,theofilis2011global} and \citet{juniper2014modal}.

The matrix-forming method was applied in the global analysis due to its flexibility and direct insight into the physical mechanism of the eigenvalues and eigenfunctions. The matrix storage and inversion issues associated with the matrix-forming method were overcome by applying the ARPACK algorithm (\citet{lehoucq1998arpack}), which utilizes the implicitly restarted Arnoldi method (IRAM) to solve the eigenvalue problem of the sparse matrices. The linearized operators were discretized on a structured curvilinear C-grid using second-order central finite difference discretization, with one-sided differences on the wall ($\Gamma_\text{w}$), far-field ($\Gamma_\infty$) and outflow ($\Gamma_\text{o}$) boundaries, as illustrated in figure~\ref{fig:grid-cylinder}. The construction of the derivative matrices, which incorporates the curvilinear coordinate transformation, is detailed in \ref{app:metrics}. The base flow coefficients in eq.~\eqref{eq:matrix-form} were obtained by direct numerical simulation~(DNS) discussed in \S\ref{sec:SU2}. 

\subsection{DNS: Base Flow}
\label{sec:SU2}

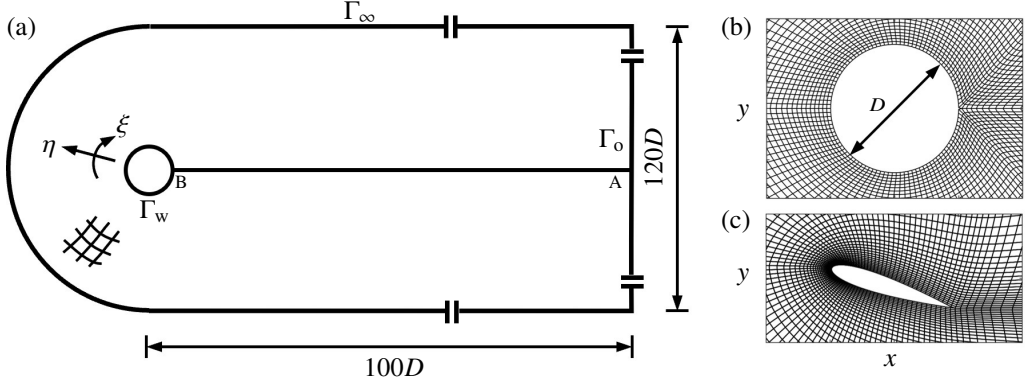
\begin{figure}
	\centering
	\begin{overpic}[width=1.0\linewidth]{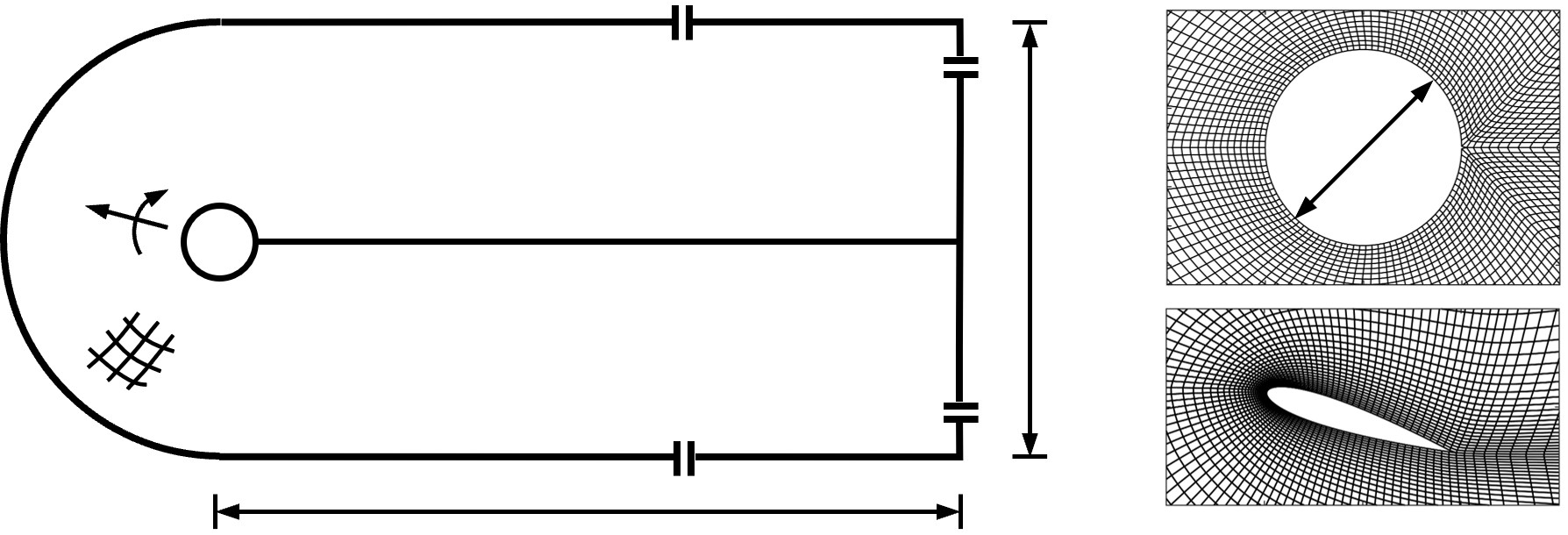}
		\put(0, 32){(a)}
		\put(70, 32){(b)}
		\put(70, 13){(c)}
		\put(33, 33.5){$ \Gamma_\infty $}
		\put(13, 14){$ \Gamma_\text{w} $}
		\put(58, 21){$ \Gamma_\text{o} $}

		\put(3.5, 20.5){$ \eta $}
		\put(10.8, 22.5){$ \xi $}
		
		\put(59,17){{\scriptsize A}}
		\put(16.5,17){\scriptsize B}
		
		\put(35, -1.5){$ 100D $}
		\put(62.5, 17){\rotatebox{90}{$120D $}}
		
		\put(84.5, 24.5){\scriptsize $ D $}
		
		\put(86, -0.5){$ x $}
		\put(71.5, 24){$ y $}
		\put(71.5, 8){$ y $}
		
	\end{overpic}
	\caption{(a)~Schematic of the computational domain for a flow past a circular cylinder of diameter $D$, as indicated in~(b). The domain spans $100D$ in the streamwise direction and $60D$ in the transverse direction on either side of the wake cut AB. Computational coordinates are represented by $\xi$ and $\eta$. Boundary conditions include the solid wall ($\Gamma_\text{w}$), far-field ($\Gamma_\infty$), and outlet ($\Gamma_\text{o}$). Panel~(b) illustrates the grid near the cylinder (displaying every 6 grid points). Panel~(c) illustrates the grid near the NACA0015 airfoil (displaying every 6 grid points). Freestream direction is from left to right along the $x$ axis.} 
	\label{fig:grid-cylinder}
\end{figure}

The computational domain, illustrated in figure~\ref{fig:grid-cylinder}, was discretized by a structured C-type curvilinear grid generated in Pointwise v18.1. The spatial domain extends 100$D$ downstream of the cylinder, where $D$ denotes the cylinder diameter that is used to normalize all spatial variables. In the transverse direction, the domain spans 60$D$ on either side of the wake cut AB. The curvilinear coordinate system comprises $1961 \times 401$ grid points along the computational coordinates $\xi$ and $\eta$, respectively, which are mapped onto the physical coordinates along the streamwise ($x$) and transverse ($y$) directions.

The cylinder flow DNS was conducted by solving the incompressible Navier-Stokes equations using SU2, an open-source multiphysics simulation and design suite (\citet{economon2016su2}). The Reynolds number, defined as $\R = U_\infty D/\nu$, was defined based on the freestream velocity $ U_\infty $, cylinder diameter, and kinematic viscosity $\nu$. The convergence of the base flow was confirmed by ensuring that the residuals reached machine precision. For the base flow simulation, the no-slip and no-penetration boundary conditions ($U = V = 0$) were imposed on the velocity components at the solid wall. At the far-field boundary, a uniform streamwise velocity and a Dirichlet condition for the transverse velocity ($U = 1$, $V = 0$) were applied. At the outlet, a Dirichlet BC was specified for the pressure ($P = 0$) and a Neumann BC for the velocity components ($\p U/\p x = \p V/\p x = 0$). The base flow velocities were then substituted into the LNSE for the eigenvalue decomposition.

\subsection{Perturbation Boundary Conditions in Wake Flows}
\label{subsec:BC}

The boundary conditions implemented in previous studies of global instabilities of two-dimensional wake flows are summarized in Table~\ref{tab:BC-literature}. The no-slip, no-penetration condition -- implemented via Dirichlet boundary conditions -- was consistently implemented at the solid wall across all referenced works. At the inlet/far-field boundaries, Dirichlet conditions were also commonly implemented, although some studies adopted an alternative Neumann condition (\citet{he2017linear}) or symmetry boundary conditions (\citet{canuto2015two}). Accordingly, we imposed Dirichlet BC on the velocity components at both the solid wall and the far-field boundaries. For the pressure mode at the wall, the linearized pressure Poisson equation (LPPE) was applied following \citet{theofilis2017linearized}.

\begin{table}
	\centering
	\setlength{\tabcolsep}{1.5pt} 
	\begin{tabularx}{\textwidth}{c *{9}{>{\centering\arraybackslash}X}}
	\hline
		& \multicolumn{3}{c}{Wall} & \multicolumn{3}{c}{Far-Field} & \multicolumn{3}{c}{Outlet} \\
		\cmidrule(lr){2-4} \cmidrule(lr){5-7} \cmidrule(lr){8-10}
		& $\hat{u}$ & $\hat{v}$ & \multicolumn{1}{c}{ $\hat{p} $} & $\hat{u}$ & $\hat{v}$ & \multicolumn{1}{c}{$\hat{p}$}& $\hat{u}$ & $\hat{v}$ & $\hat{p}$ \\ \hline \addlinespace
		\citet{barkley1996three}			& D & D & - & D & D & - & N & N & D \\ 
		\citet{giannetti2007structural}		& D & D & - & N (D) & D & N & N / - & D / N & D / N \\ 
		\citet{he2017linear}				& D & D & - & N (D) & N (D) & - & N & N & D \\ 
		\citet{mittal2010stability}			& D & D & - & $\sigma_{xy}$ & D & - & $\sigma_{xx}$ & $\sigma_{xy}$ & - \\ 	
		\citet{sipp2007global}				& D & D & - & N (D) & D & - & $\sigma_{xx}$ & N & - \\
		\citet{canuto2015two}				& D & D & - & S & S & S & SL & SL & SL \\
		\citet{theofilis2003advances}		& D & D & - & D & D & D & E & E & E \\
		Present study						& D & D & LPPE & D & D & D & \multicolumn{3}{c}{N / D / E / R / SL / $\sigma_{ij}$} \\
		\hline
	\end{tabularx}
	\caption{Perturbation BCs for global stability analysis of wake flows past bluff bodies. N-Neumann; D-Dirichlet; R-Robin; S-symmetry; SL-sponge layer; $\sigma_{ij}$-stress tensor; E-extrapolation; LPPE-linearized pressure Poisson equation. (Parenthesis in Far-Field denote inlet BC if different from top/bottom far-fields)}
	\label{tab:BC-literature}
\end{table}

In contrast to the general consensus on wall and inlet/far-field boundary conditions, there was significantly less agreement regarding the appropriate treatment at the outlet of the computational domain. As noted by \citet{sani1994resume}, an effective OBC should permit flow structures to convect downstream without generating divergence errors or inducing spurious upstream reflections — an aspect of particular importance in wake flows where the disturbance can continue to grow spatially past the outlet. Therefore, the focus of this study is to analyze the impact of different OBCs on the global stability analysis of the wake flow. 

\subsubsection{Non-physical Outflow Boundary Conditions}

The main OBCs from the literature include: 1) Classical Neumann or Dirichlet conditions applied separately to the velocity and pressure modes (\citet{barkley1996three, giannetti2007structural, he2017linear}); 2) Stress-free boundary conditions, in which the stress tensor was defined as: $ \sigma_{ij} = -\hat{p}\delta_{ij} + \R^{-1} \left(\p \hat{u}_i/\p x_j + \p \hat{u}_j/\p x_i \right)$ implemented in \citet{noack2003hierarchy, mittal2010stability, sipp2007global}; and 3) Other methods such as sponge layers designed to mitigate contamination due to inaccurate or non-reflecting outlet conditions (\citet{canuto2015two}), with detailed guidance on their implementation provided in \citet{mani2012analysis} and the extrapolation method discussed in \citet{theofilis2003advances, chedevergne2006biglobal}. These classical OBCs lack the capability to accurately capture the true physics of the instability modes at the outflow. 

\subsubsection{Physical Robin Boundary Condition: Incorporating Local LST}
\label{subsec:Robin}

A natural way to reveal the spatial information is to incorporate the local linear stability theory (LST) at the outlet to the OBC of the global analysis. The concept of linking temporal and spatial growth rates originates from~\citet{gaster1962note}, who showed that the real and imaginary components of the group velocity satisfy the Cauchy–Riemann equations,

\be\label{eq:CR}
c \equiv \f{\p \omega}{\p \alpha} = c_r + \ti c_i,\quad
c_r = \f{\p \omega_r}{\p \alpha_r} = \f{\p \omega_i}{\p \alpha_i},\quad
c_i = \f{\p \omega_i}{\p \alpha_r} = -\f{\p \omega_r}{\p \alpha_i},
\ee
where $c$ is the group velocity and $\alpha$ represents the streamwise wavenumber, formally defined by the local LST ansatz

\be\label{eq:local}
\tilde{\bq} = \hathat{\bq}(y)e^{\ti\left(\alpha x + \beta z - \omega t\right)},
\ee
where $\hathat{\bq}$ is the local eigenfunction. The relations between $\omega$, $\alpha$ and $\hathat{\bq}$ are given by the Orr-Sommerfeld and Squire equations (see, e.g.,~\cite{juniper2014modal,SchmidHenningson2001,theofilis2011global}).
The Gaster transform is based on the integration of eq.~\eqref{eq:CR} and a series expansion of $c_r$, which yields

\be\label{eq:Gaster}
\omega_i = -c_r\alpha_i,
\ee
relying on the imaginary part of the group velocity, $c_i$, being rather small for boundary layer and channel flows. The real part of the group velocity, $c_r$, can be evaluated at any station between the temporal case ($\alpha\in\mathbb{R}$, $\omega\in\mathbb{C}$) and the spatial case ($\omega\in\mathbb{R}$, $\alpha\in\mathbb{C}$).

\citet{ehrenstein2005two} and \citet{alizard2007spatially} used a similar analysis based on a Taylor series around the neutral point, $(\omega_0,\alpha_0)\in\mathbb{R}$,

\be\label{eq:Taylor1}
\omega - \omega_0 = c_0(\alpha - \alpha_0),
\ee
where the subscript `0' indicates the neutral point. The real and imaginary parts of eq.~\eqref{eq:Taylor1} are

\bse\label{eq:Taylor2}
\be\label{eq:wr_exact}
\omega_r - \omega_0 = c_{0r}(\alpha_r - \alpha_0) - c_{0i}\alpha_i,
\ee
\be\label{eq:wi_exact}
\omega_i = c_{0r}\alpha_i + c_{0i}(\alpha_r - \alpha_0).
\ee
\ese
Similarly to \citet{gaster1962note}, \citet{ehrenstein2005two} and \citet{alizard2007spatially} assumed that the imaginary part of the group velocity, $c_{0i}$, is small, as their focus was on boundary layers. Therefore, neglecting $c_{0i}$, eq.~\eqref{eq:Taylor1} becomes

\be
\label{eq:Taylor3}
\omega - \omega_0 = c_{0r}(\alpha - \alpha_0),
\ee
with real and imaginary parts

\bse\label{eq:Taylor4}
\be\label{eq:wr_approx}
\omega_r - \omega_0 = c_{0r}(\alpha_r - \alpha_0),
\ee
\be\label{eq:wi_approx}
\omega_i = c_{0r}\alpha_i.
\ee
\ese
It should be noted that eq.~\eqref{eq:wi_approx} does not contradict eq.~\eqref{eq:Gaster} since Gaster's derivation is not conducted around the neutral point; notwithstanding, it can be recovered by the following substitution in eq.~\eqref{eq:Taylor1}: the temporal case is plugged into ($\omega,\alpha$) and the spatial case into ($\omega_0,\alpha_0$), i.e. $\omega=\omega_r+\ti\omega_i$, $\alpha=\alpha_r$, and $\omega_0=\omega_r$, $\alpha_0=\alpha_r+\ti\alpha_i$, respectively. The underlying conclusion of both eq.~\eqref{eq:Gaster} and eq.~\eqref{eq:wi_approx} is the same -- the former enables converting temporal growth rate into spatial growth rate, whereas the latter states that the temporal growth rate of a global eigenmode comes at the expense of its local spatial growth rate.

An interesting conclusion arises from the above analysis with regard to the spatial growth of the global eigenmode very far downstream, where the wake has straightened out and the local velocity profile is uniform, i.e. $U(x\to\infty, y) = U_c$, where $U_c$ is constant. For that case, an analytical solution for the local eigenmodes and the group velocity exists, given by

\be
\omega=\alpha U_c - \f{\ti k^2}{\R} \approx \alpha U_c,\quad c_0\equiv\f{\p\omega}{\p\alpha}\bigg|_0 = U_c -\f{2 \ti\alpha_0}{\R} \approx U_c, \label{eq:U-straight}
\ee
where $k^2$ is the total spatial wavenumber (corresponding to the Fourier transform of $-\nabla^2$). In the globally stable (subcritical) case, where $\R<\R_{cr}$, the temporal growth rate is negative, i.e., $\omega_i<0$. Using eq.~\eqref{eq:wi_approx} it can be seen that $-\alpha_i\approx-\omega_i/U_c>0$, stating that the global eigenfunction is expected to amplify spatially {\textit{ad infinitum}}, even as $x\to\infty$. This may preclude convergence of the global stability eigenmodes unless a proper OBC is applied to capture its behavior; moreover, further extension of the domain will not facilitate convergence.

The Robin boundary condition is derived similarly to \citet{ehrenstein2005two} and \citet{alizard2007spatially} by applying the Fourier transform of the streamwise derivative, $\ti\alpha\hat{q}=\p\hat{q}/\p x$, which yields

\be
\ti \left( \omega_0 - \alpha_0 c_0 \right) \hat{\bs{q}} + c_0 \f{\p \hat{\bs{q}}}{\p x} = \ti\omega \hat{\bs{q}} \label{eq:Robin1},
\ee
noting that this boundary condition is imposed without knowing $\omega$ a priori, which is part of the solution of the eigenvalue decomposition system. Two versions of the boundary condition are tested:
the Robin-R assumes that $c_{0i}$ is negligible (based on eq.~\eqref{eq:Taylor3}), whereas the Robin-C is derived for the general case where $c_0\in\mathbb{C}$ (based on eq.~\eqref{eq:Taylor1}). The differences between both versions of the OBC will be quantified in detail in \S\ref{sec:results}. It should be noted that if we assume $\omega_0 = \alpha_0 c_0$ in eq.~\eqref{eq:Robin1}, the current Robin condition becomes the classical convective boundary condition (e.g.,~\citep{orlanski1976simple}), except that the value of $c_0$ in eq.~\eqref{eq:Robin1} is determined from local linear stability theory, rather than relying on a prescribed convection velocity $U_c$. Therefore, Robin-R serves as a representative example of the classical convective boundary condition.

The group velocity was calculated based on the biorthogonality of the spectrum (e.g.,~\citet{Nayfeh1979,Tumin2011}), utilizing the adjoint eigenvector,

\be\label{eq:delta_alpha}
c_0=\f{\p\omega}{\p\alpha}\bigg|_0 = \f{\left<\hathat{\bm{q}}^{\dagger},\mathcal{N}\hathat{\bm{q}}\right>}{\left<\hathat{\bm{q}}^{\dagger},\mathcal{M}\hathat{\bm{q}}\right>}, 
\ee
with the inner products given by

\bse
\be
\left<\hathat{\bm{q}}^{\dagger},\mathcal{N}\hathat{\bm{q}}\right>=
\int_{-\infty}^{\infty} \biggl( \left(\overline{U}(\xo,y) - \f{2\ti\alpha_0}{\R}\right) \left( \hathat{u}^{\dagger}\hathat{u} - \hathat{v}^{\dagger}\hathat{v} + \hathat{w}^{\dagger}\hathat{w} \right) + \hathat{p}^{\dagger}\hathat{u} + \hathat{u}^{\dagger}\hathat{p} \biggr)\;\td y,
\ee
\be
\left<\hathat{\bm{q}}^{\dagger},\mathcal{M}\hathat{\bm{q}}\right>=
\int_{-\infty}^{\infty} \left( \hathat{u}^{\dagger}\hathat{u} - \hathat{v}^{\dagger}\hathat{v} + \hathat{w}^{\dagger}\hathat{w} \right)\;\td y,
\ee
\ese
where $\hathat{\bs{q}}$ and $\hathat{\bs{q}}^{\dagger}$ are the eigenvectors of the direct and adjoint local LST, respectively, and $\overline{U}(\xo,y)$ is the base flow wake profile at the outlet, shown for selected values of $\xo$ in \ref{app:wake-profile}.

\section{Results and Discussions}
\label{sec:results}

In this section, we first show the impact of OBCs on global stability analysis of cylinder wakes at different $\R$, ranging from the subcritical (stable) to the unstable regime. Both the eigenvalues and the eigenfunctions are compared for different OBCs, with special attention given to the pressure component.  Unless otherwise specified, all outflow boundary conditions (OBCs)---namely Neumann, Dirichlet, extrapolation, and two Robin conditions---are applied consistently to both velocity and pressure components. The same analysis is then conducted for the airfoil wake for both two- and three-dimensional perturbations, to demonstrate the applicability of the present conclusions for an arbitrary wake around a bluff body. 

\subsection{Cylinder Wake in the Globally Stable Regime}
\label{subsec:cylinder-low-Re}

\subsubsection{Neumann, Dirichlet, Extrapolation, and Stress-free Conditions}
\label{subsec:Neumann-Dirichlet}

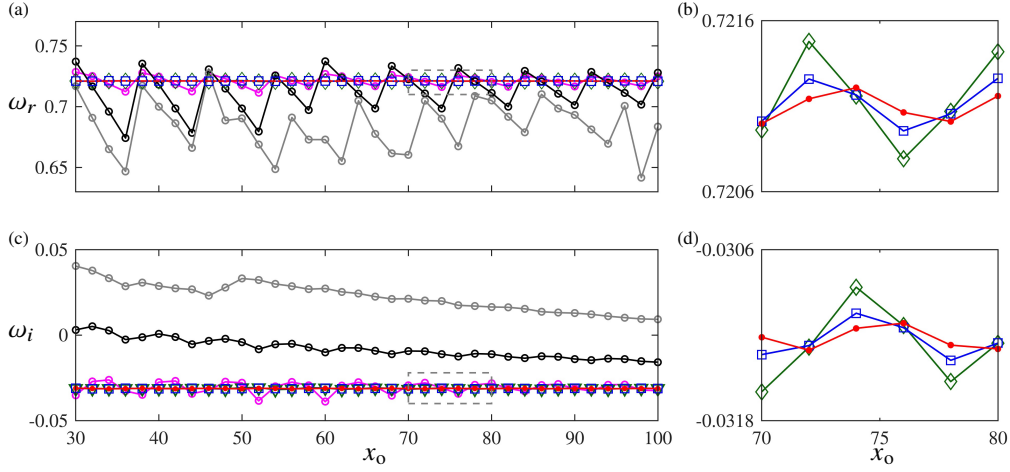
\begin{figure}
	\centering
	\begin{overpic}[width=0.633\linewidth]{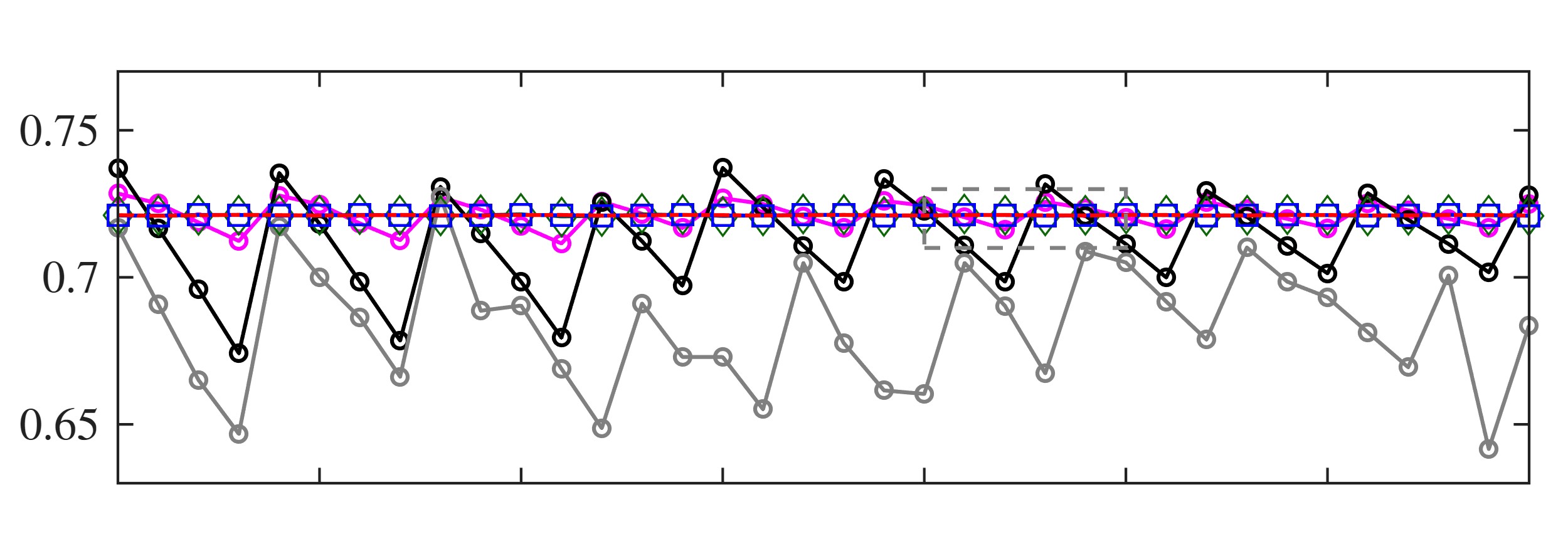}
		\put(-3, 31.5){\scriptsize(a)}
		\put(-3, 17){$ \omega_r $}
	\end{overpic}
    \begin{overpic}[width=0.3167\linewidth]{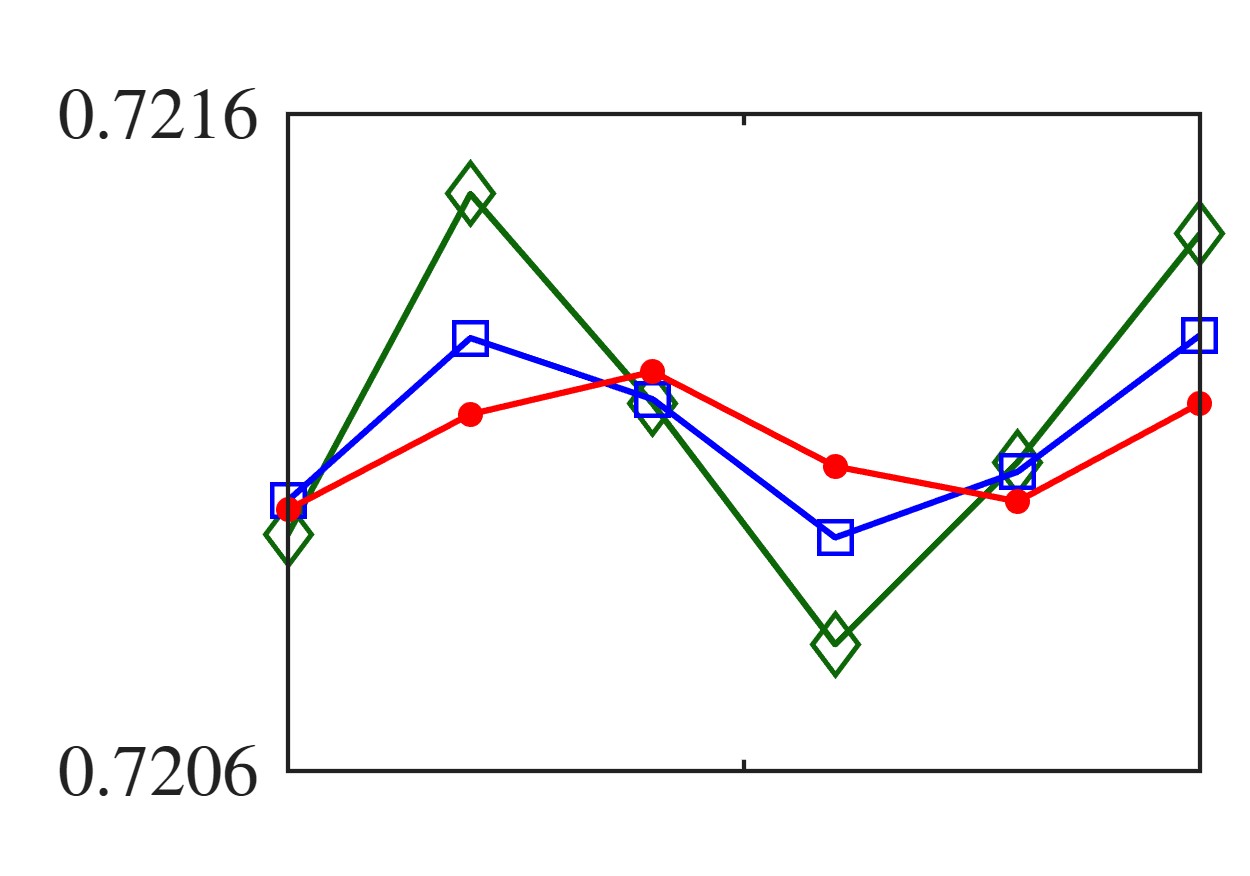}
		\put(-3, 63){\scriptsize(b)}
	\end{overpic}
    \begin{overpic}[width=0.633\linewidth]{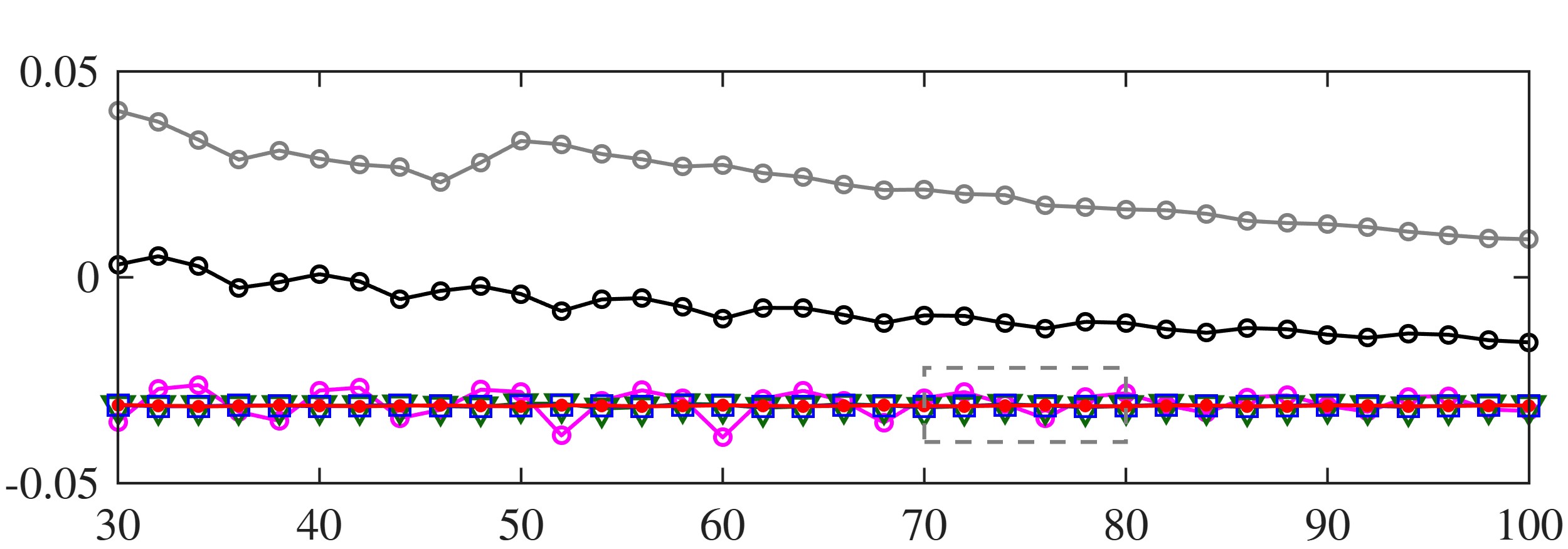}
		\put(-3, 31.5){\scriptsize(c)}
		\put(-3, 17){$ \omega_i $}
		\put(52,  -2){$ x_\text{o} $}
	\end{overpic}
    \begin{overpic}[width=0.3167\linewidth]{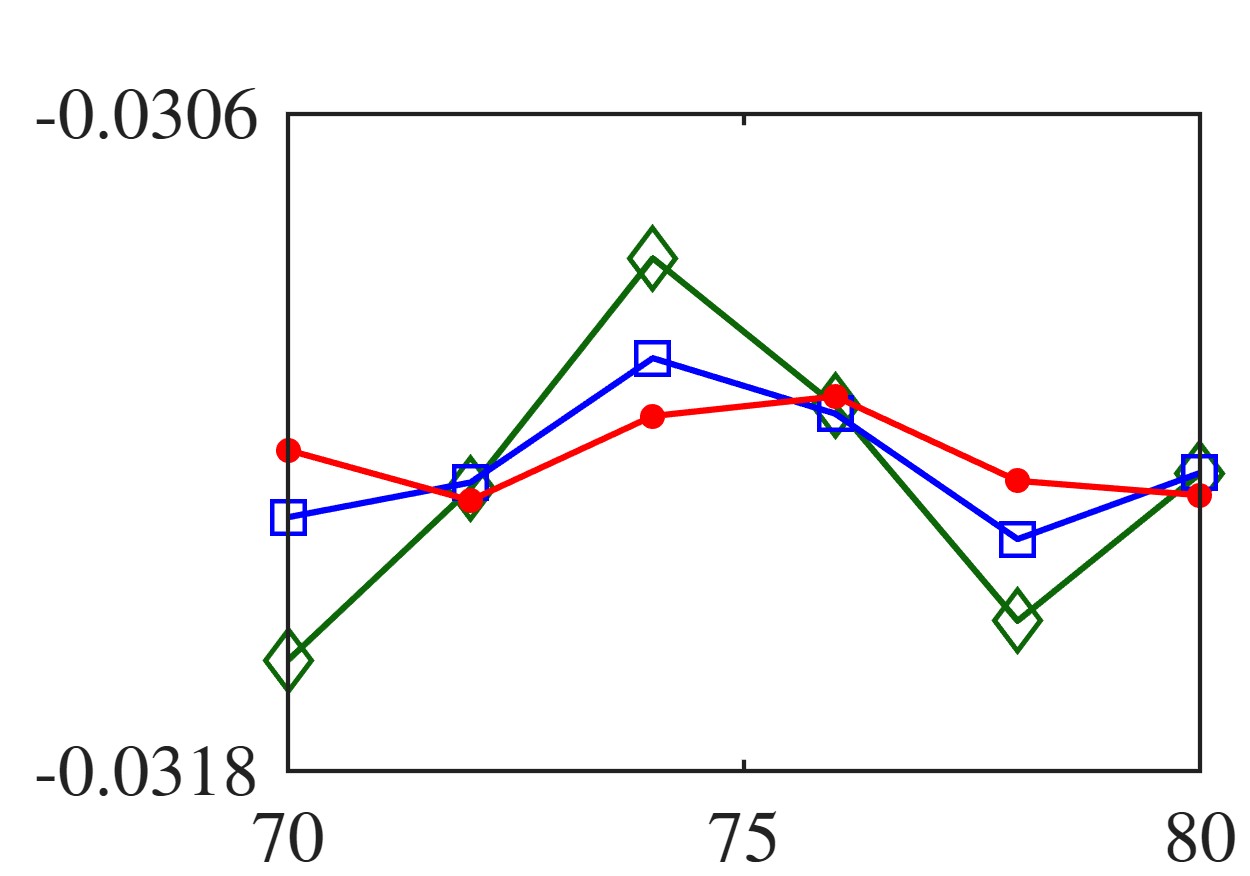}
		\put(-3, 63){\scriptsize(d)}
		\put(55,-4){ $ x_\text{o} $ }
	\end{overpic}
\caption{Real~(a) and imaginary~(c) parts of the least stable eigenvalue from the global stability analysis for different outlet locations, $\xo$, of the cylinder wake at $ \R = 40 $. Different outflow boundary conditions are denoted by different colors: Dirichlet (magenta), Neumann (black), Extrapolation (gray), Stress-free (green), Robin-R (blue), and Robin-C (red). Panels (b, d) provide magnified views of the regions indicated by the gray dashed boxes in (a, c).}
	\label{fig:spectrum-different-BCs-cylinder-40}
\end{figure}

Figure~\ref{fig:spectrum-different-BCs-cylinder-40} shows the eigenvalue of the least stable global mode (the mode with maximum $\omega_i$, following \citep{SchmidHenningson2001}) for the cylinder wake at $\R=40$, which is lower than the critical $\R_{cr} \approx $ 47. To assess the spatial convergence of the spectrum, the eigenvalue problem was solved for a wide range of domain truncations, with outlet locations varying as $ 30 \leq \xo \leq 100 $, where $\xo$ represents the outlet location in the global analysis. The least stable eigenmode for the Neumann BC (black) exhibits high sensitivity to $\xo$ and does not converge well even for the case of the maximum domain size at $\xo = 100$. The Dirichlet BC (magenta) significantly improves the convergence of the eigenvalue with $\xo$, although the variation at large $\xo$ remains. The present results suggest that the Neumann OBC tends to cause inaccurate $\omega_r$ and $\omega_i$ compared to the Dirichlet condition. Such behavior is observed for the eigenfunctions as well (discussed later). The linear extrapolation method (gray) exhibits even higher variation in $ \omega_r $ with domain size, as well as incorrect $ \omega_i $ (in both current subcritical and supercritical $\R$ discussed in \S\ref{subsec:Re-effect-cylinder}), and therefore the results of the extrapolation OBC are excluded in subsequent discussions.

Figure~\ref{fig:spectrum-different-BCs-cylinder-40} reveals that in the subcritical (stable) case, global stability analysis is highly sensitive to OBCs, and classical Neumann and Dirichlet conditions fail to yield the converged spectrum. Notably, this issue cannot be overcome simply by extending the computational domain in the wake, which is perhaps challenging the general assumption that the impact of OBCs in DNS can be alleviated by employing a larger computational domain, with its size scaling linearly with Reynolds numbers (\citet{dong2014robust}). However, figure~\ref{fig:spectrum-different-BCs-cylinder-40} demonstrates that the variation in the eigenmode for the Neumann and Dirichlet conditions does not diminish with increasing $\xo$. This suggests that the linear scaling of the computational domain with $\R$ really depends on the specific problem, and simply extending the domain without implementing the appropriate OBC cannot guarantee convergence in the global stability analysis. The stress-free boundary condition (following~\citet{mittal2010stability}), indicated by the green color, shows a significant convergence of the spectrum compared to the Neumann and Dirichlet conditions, with a magnified view of its variation with $\xo$ shown in figure~\ref{fig:spectrum-different-BCs-cylinder-40}(b, d).

Figure~\ref{fig:u-mode-cylinder-Re-40} compares the least stable eigenmodes for different OBCs reported in figure~\ref{fig:spectrum-different-BCs-cylinder-40}, with the outlet positioned at $\xo = 52$ (representative of the behavior for other $\xo$). The velocity components for the Neumann (a, b) and Dirichlet (d, e) BC exhibit similar behavior, though the Neumann condition yields higher distortion near the outlet, consistent with the higher oscillation of the eigenvalue in figure~\ref{fig:spectrum-different-BCs-cylinder-40}. The pressure component for these classical boundary conditions differs significantly: the $\hat{p}$ mode from the Neumann BC (c) displays a huge distortion compared to the Dirichlet BC (f). This distortion is not confined to only very close to the outlet, but to almost the entire computational domain. These findings are consistent with the unconverged eigenvalues of the Neumann OBC, indicating that the Neumann OBC tends to cause distortions over a wide region, leading to spurious less stable modes (larger $ \omega_i $ in figure~\ref{fig:spectrum-different-BCs-cylinder-40}(b) and large distortion in figure~\ref{fig:u-mode-cylinder-Re-40}(c)). In addition, the $\hat{p}$ mode from the Dirichlet condition in (f) exhibits striped patterns that extend throughout a significant portion of the computational domain (spanning at least $10D$), rather than being confined near the outlet, as suggested by \citet{theofilis2003advances}. Similar striped distortion in $\hat{p}$ is also observed in the vicinity of the outlet of the stress-free boundary condition, shown in (i), suggesting that even though the eigenvalue is substantially converged compared to Neumann and Dirichlet conditions in figure~\ref{fig:spectrum-different-BCs-cylinder-40}, still the pressure component is not well converged. Furthermore, the effect on spectrum and eigenmodes by imposing a sponge layer at the outlet is discussed in~\ref{app:convect-sponge}.

The comparison between the velocity and pressure components shows that the latter is more sensitive to the choice of OBCs in the global stability analysis (\citet{tomboulides1993direct}) and therefore essential to evaluate the effectiveness of OBCs. Despite the significance of pressure modes in evaluating the quality of global results, it seems somewhat surprising that most previous studies only showed the velocity modes, and only a few studies showed the pressure mode (e.g.,~\citet{mittal2010stability}). 

The elevated noise levels associated with the Neumann, Dirichlet, and stress-free OBCs can be attributed to the characteristics of the eigenmodes, as shown in figure~\ref{fig:u-mode-cylinder-Re-40}. The global modes display spatial growth in the streamwise direction, which is incompatible with the assumption of Neumann or Dirichlet conditions, which presume either zero spatial growth or zero amplitude at the outlet, respectively; similarly, the simple assumption of zero stress at the outlet can not perfectly capture the true spatial behavior of the pressure modes. Consequently, all these conditions cannot accurately capture the spatially growing nature of local disturbances and are not suitable as physically representative OBCs. A more appropriate OBC, aligned with the spatial growth of the eigenmodes, is therefore essential to capture the true physics at the outlet.

\subsection{Robin Outflow Boundary Conditions}
The blue and red lines in figure~\ref{fig:spectrum-different-BCs-cylinder-40} show the eigenvalue of the two Robin OBCs discussed in \S\ref{subsec:Robin}, denoted by Robin-R and Robin-C. The eigenvalues of the two Robin conditions have a significant improvement in convergence compared to the classical Neumann and Dirichlet BC. For the range of $\xo$ considered in this study, the variation of $ \omega_i $ for the Dirichlet BC is about 2\%, which is similar to the uncertainty level reported in Table~7 in \citet{he2017linear}. The variations in $\omega_i$ are 0.4\% and 0.2\% for the Robin-R and Robin-C, respectively. This demonstrates that the Robin boundary condition ensures a better converged spectrum irrespective of the outlet position of the computational domain.

The quality of the pressure component can be quantified by a relative error,

\be
    \varepsilon_{\hat{p}}  = \frac{\| \hat{p}_\text{o} - \hat{p}_\text{o}^\text{ref} \|_2}{ \| \hat{p}_\text{o}^\text{ref} \|_2},
    \label{eq:gamma-p}
\ee
where $\| \cdot \|_2$ denotes the $L_2$ norm, $\hat{p}_\text{o}$ represents the pressure mode at the outlet for different OBCs, and $\hat{p}_\text{o}^\text{ref}$ denotes the pressure mode corresponding to the Robin-C condition, which is defined as a reference. All $\hat{p}_\text{o}$ are normalized to be identical at $(x,y) = (x_\text{o}, 2)$. The value of $\varepsilon_{\hat{p}}$ for the Dirichlet condition should be unity due to its nature (as $\hat{p}_\text{o} = 0$ at the outlet). The values of $\varepsilon_{\hat{p}}$ for the Neumann and stress-free conditions are 1.7 and 0.17, respectively. The Robin-R condition has $\varepsilon_{\hat{p}} \approx 0.06$. The value of $\varepsilon_{\hat{p}}$ quantifies the relative error in the pressure component, observed in the contour in figure~\ref{fig:u-mode-cylinder-Re-40}, among different boundary conditions.

\begin{landscape}
\begin{figure}
	\begin{overpic}[height=15cm]{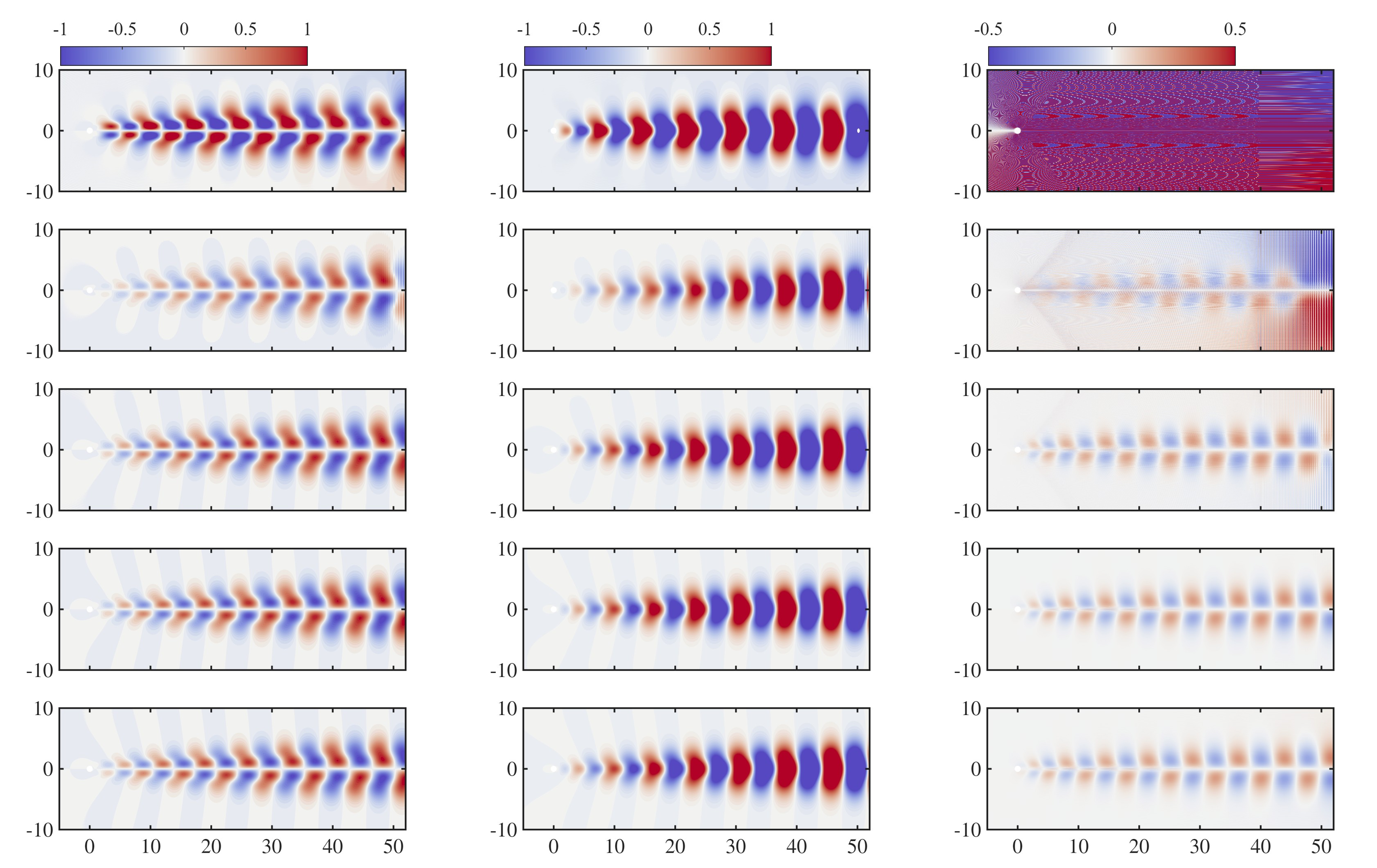}
		\put(0, 58){(a)}
		\put(33.6, 58){(b)}
		\put(67, 58){(c)}
		\put(0, 46){(d)}
		\put(33.6, 46){(e)}
		\put(67, 46){(f)}
		\put(0, 34.2){(g)}
		\put(33.6, 34.2){(h)}
		\put(67, 34.2){(i)}
		\put(0, 22.8){(j)}
		\put(33.6, 22.8){(k)}
		\put(67, 22.8){(l)} 
        \put(-0.5, 11){(m)}
		\put(33.6, 11){(n)}
		\put(67, 11){(o)} 
        
		\put(23.5, 58.5){$\Re(\hat{u})$}
		\put(57.5, 58.5){$\Re(\hat{v})$}
		\put(91, 58.5){$\Re(\hat{p})$} 

		\put(17, -0.5){$x$}
		\put(51, -0.5){$x$}
		\put(84.3, -0.5){$x$}
		
		\put(1, 53.3){$y$}
		\put(1, 41.7){$y$}
		\put(1, 30){$y$}
		\put(1, 18.4){$y$}
        \put(1, 7){$y$}
	\end{overpic}
	\caption{Real part of the least stable global eigenmode of the flow past a circular cylinder at $\R=40$: $ \hat{u} $ (first column), $ \hat{v} $ (second column), and $ \hat{p} $ (third column), for different outflow boundary conditions: Neumann (a,b,c), Dirichlet (d,e,f), Stress-free (g,h,i), Robin-R (j,k,l), and Robin-C (m,n,o). The outlet is located at $\xo = 52$. All eigenfunctions are normalized by the magnitude of the $ \hat{u} $ mode at $(x,y)\approx (48, 2)$. The relative error, defined in eq.~\eqref{eq:gamma-p}, is $ \varepsilon_{\hat{p}}$ = 1.74 (c), 1 (f), 0.17 (i), 0.06 (l), and 0 (o).}
	\label{fig:u-mode-cylinder-Re-40}
\end{figure}
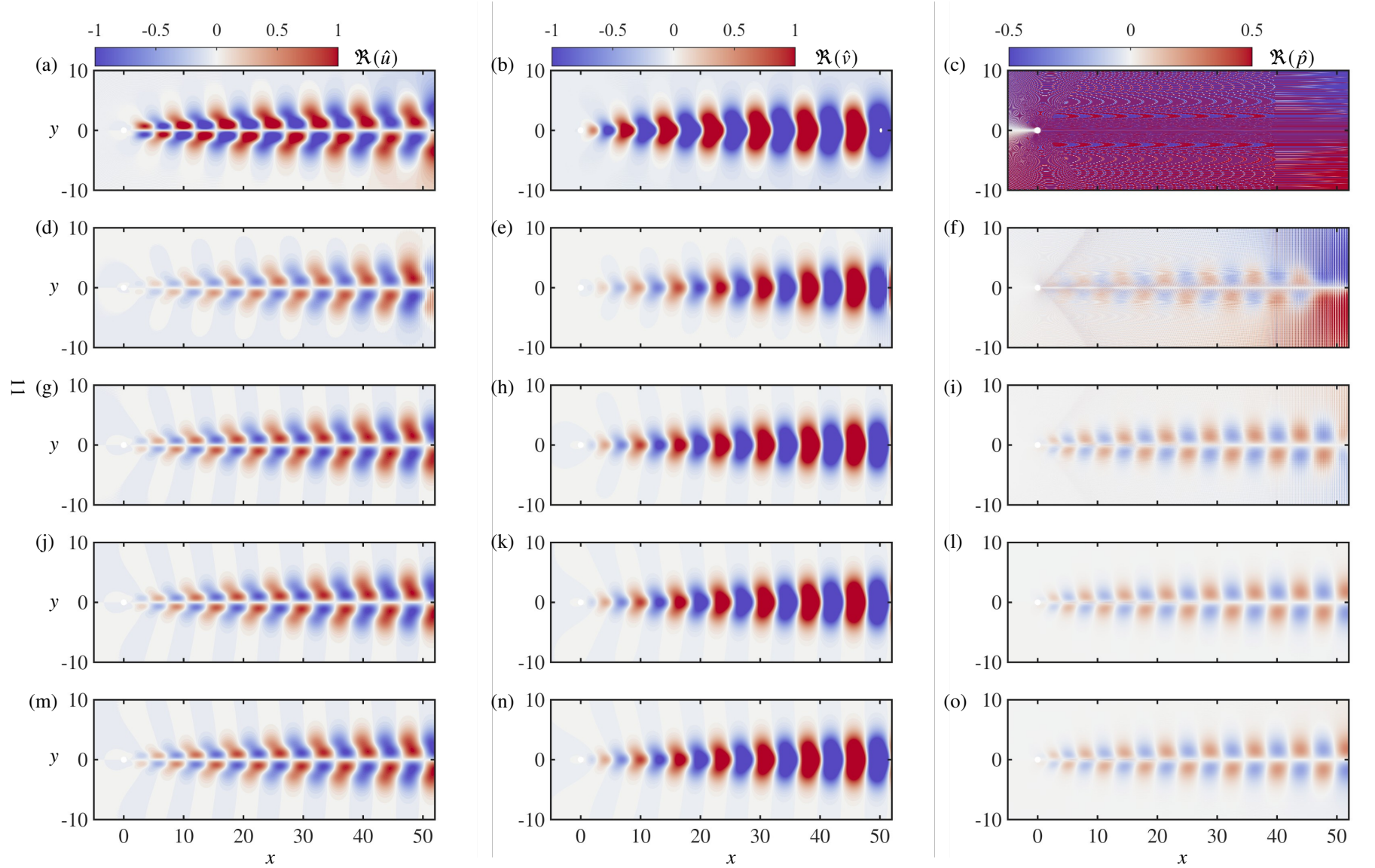
\end{landscape}

The significant improvement associated with the Robin BC is observed in the eigenmodes in figure~\ref{fig:u-mode-cylinder-Re-40}, particularly for the pressure components. Figure~\ref{fig:u-mode-cylinder-Re-40}(l, o) illustrate $\hat{p}$ for both Robin BCs. The high level of noise was eliminated by imposing the Robin BC, even very close to the outlet. The better performance of the eigenfunction does not depend on the choice of outlet positions, and the difference between the two Robin conditions seems to be negligible, indicating that qualitatively, whether or not including the imaginary part in the group velocity does not seem to change the global results. In other words, it seems safe to neglect $c_{0i}$ in the current wake flow study like in the boundary layer studies \cite{alizard2007spatially}. A critical assessment of the difference between the Robin-R and Robin-C conditions is conducted in the following section.

\subsubsection{Real vs. Complex Robin Condition}
\label{subsec:real-compelx-Robin}

\begin{figure}
	\centering
	\begin{overpic}[width=0.9\linewidth]{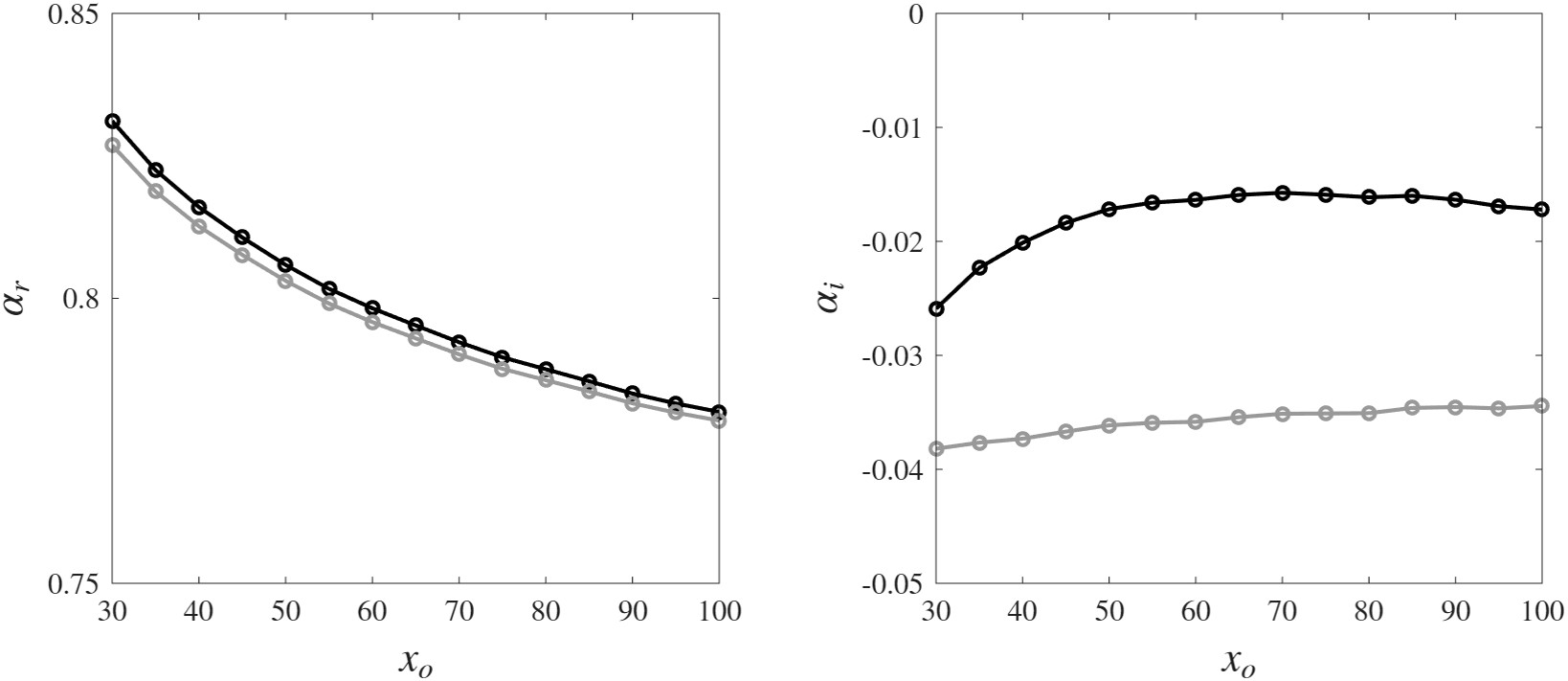}
		\put(-1, 43){(a)}
		\put(51, 43){(b)}
	\end{overpic}
\caption{Real~(a) and imaginary~(b) parts of $\alpha$ estimated in eq.~\eqref{eq:Taylor2} and eq.~\eqref{eq:Taylor4}, for Robin-R condition (gray) and Robin-C condition (black). The eigenvalue $\omega$ was obtained from global analysis. } 
	\label{fig:alpha-actual-comparison-Taylor-Re-40}
\end{figure}

In this subsection, the difference between the two Robin conditions is quantified by examining how well the local streamwise wavenumber evaluated from the global eigenvalue using eq.~\eqref{eq:Taylor1} matches the local spatial growth of the actual global modes. The local streamwise wavenumber can be obtained from eqs. \eqref{eq:Taylor2} and \eqref{eq:Taylor4}, for the Robin-C and Robin-R OBCs, respectively. The real and imaginary parts of the streamwise wavenumber are shown in figure~\ref{fig:alpha-actual-comparison-Taylor-Re-40}. The real part, $\alpha_r$, shown in figure~\ref{fig:alpha-actual-comparison-Taylor-Re-40}(a) is very close for the two Robin conditions throughout the entire range of $\xo$, which can be deduced by comparing the expressions in eq.~\eqref{eq:wr_exact} and eq.~\eqref{eq:wr_approx}, and this implies that neglecting $c_{0i}$ does not affect the estimation of $\alpha_r$ for both Robin conditions. However, this consistency does not hold for the imaginary part, $\alpha_i$, as can be seen in figure~\ref{fig:alpha-actual-comparison-Taylor-Re-40}(b). Ignoring the imaginary part of the group velocity, $c_{0i}$, tends to overestimate the local spatial growth rate, $-\alpha_i$, approximately by a factor of 2. Therefore, if one is interested in the local spatial growth of the global eigenmodes from the local analysis, the imaginary part of the group velocity has to be included and the Robin-C condition should be applied.

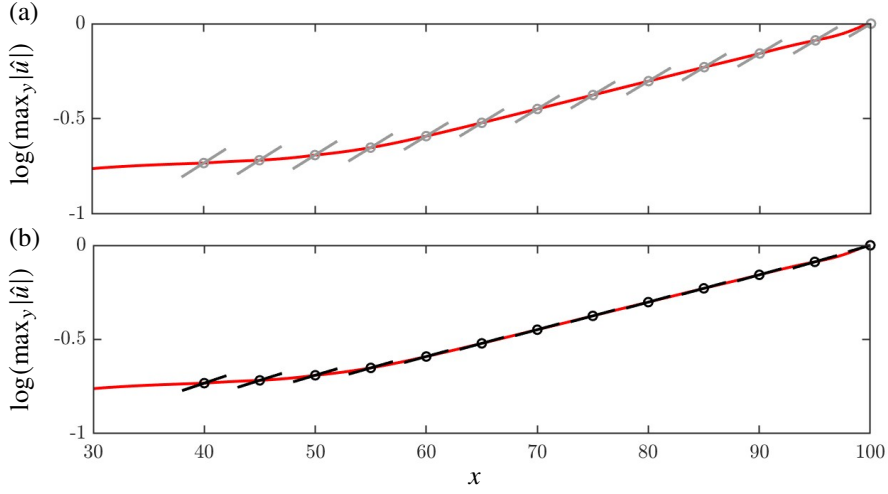
\begin{figure}
	\centering
	\begin{overpic}[width=0.85\linewidth]{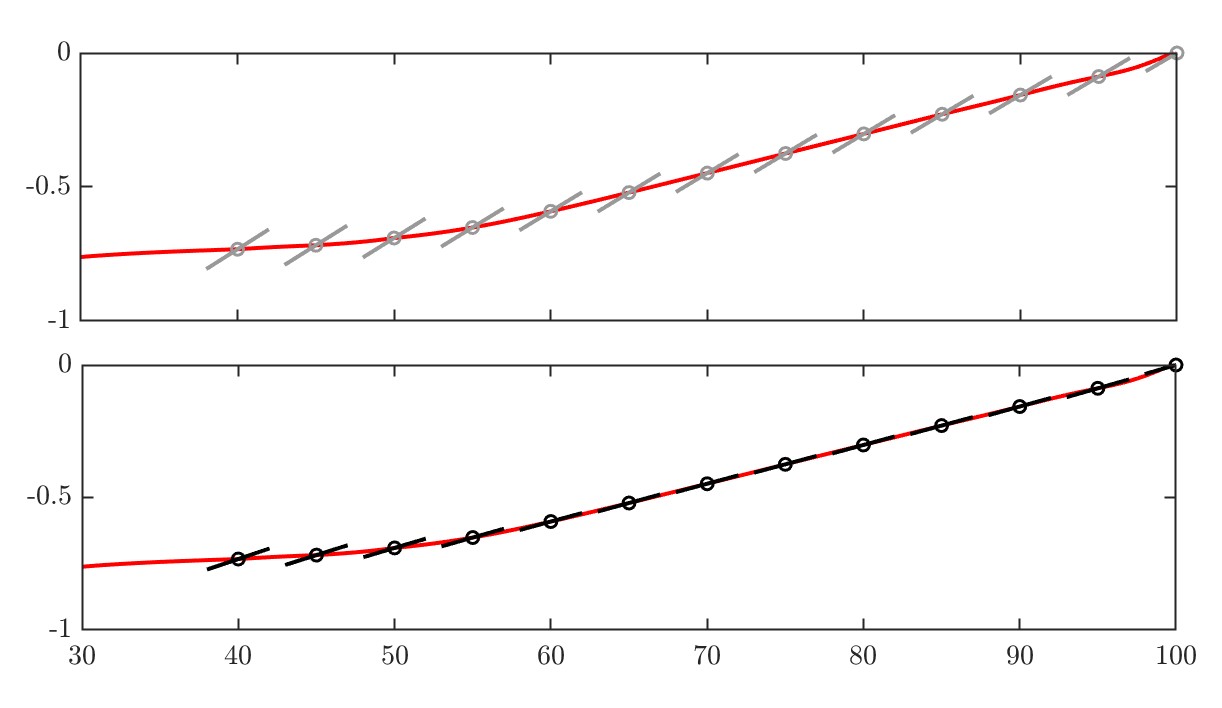}
		\put(-3, 54){(a)}
		\put(-3, 28){(b)}
		\put(50, 0.5){$ x $}	
		\put(-3, 9){\rotatebox{90}{$ \log(\text{max}_y|\hat{u}|)$}	}
		\put(-3, 35){\rotatebox{90}{$ \log(\text{max}_y|\hat{u}|)$}	}
		
	\end{overpic}
\caption{Comparison of local spatial growth rates of the streamwise velocity component of the global mode, $\hat{u}$, with local analysis. The red solid line shows the maximum magnitude of $\hat{u}$ at each streamwise position. The slope of the short lines are identical to $-\alpha_i$ evaluated at each corresponding $\xo$ location for (a)~Robin-R, based on eq.~\eqref{eq:wi_approx}; (b)~Robin-C, based on eq.~\eqref{eq:wi_exact}.}
	\label{fig:spatial-growth-rate-alphai-cylinder-Re-40}
\end{figure}

Figure~\ref{fig:spatial-growth-rate-alphai-cylinder-Re-40} illustrates the local spatial growth rate calculated from eq.~\eqref{eq:wi_approx} and eq.~\eqref{eq:wi_exact}, to compare how well the local LST captures the local spatial growth of the global eigenmodes. The red lines mark the maximum magnitude of the mode $\hat{u}$ within $30 \leq x \leq 100$ in figure~\ref{fig:u-mode-cylinder-Re-40}(m), with $ \xo = 100 $, and its slope represents the local spatial growth of the global mode. The slopes of the short lines are identical to $-\alpha_i$, evaluated at the corresponding $\xo$ locations of the local analysis, which should ideally match the slope of the red line. 

Figure~\ref{fig:spatial-growth-rate-alphai-cylinder-Re-40}(a) shows that the spatial growth rate appears to be overestimated by local analysis in the Robin-R OBC, shown by the larger slope of the short gray lines, whereas the Robin-C OBC shown in figure~\ref{fig:spatial-growth-rate-alphai-cylinder-Re-40}(b) matches much better with the spatial growth of the global mode. Therefore, even though in \S\ref{subsec:Neumann-Dirichlet} it seems that neglecting $c_{0i}$ will not have a significant impact on the spectrum and eigenfunctions in the global analysis, its contribution to $\alpha_i$ is crucial and therefore should be taken into account if the spatial growth of the global modes is of interest. In other words, the Robin-C captures more accurately the local spatial growth of the global eigenmodes, and therefore we impose the Robin-C condition when we refer to the Robin condition in the subsequent study.

\subsection{Reynolds Number Dependence of OBC Impact}
\label{subsec:Re-effect-cylinder}

In the preceding section, we examined the choice of outflow boundary conditions for the cylinder wake in the globally stable regime, showing that the spatially growing nature of the eigenmodes can lead to nonphysical behavior at the outlet when classical Dirichlet or Neumann OBCs are implemented. Because the spatial structure of the global modes depends on the Reynolds number, we now analyze how the Reynolds number affects the choice of appropriate OBC. To obtain a steady base flow for $\R \geq 47$ (and suppress vortex shedding), we solve only the upper half of the computational domain shown in figure~\ref{fig:grid-cylinder}(a), imposing a symmetry boundary condition along the centerline along the $x$ axis. The resulting half-domain solution is then mirrored to reconstruct the full flow field, as in previous studies (e.g.,~\citet{sipp2007global}).

\begin{figure}[ht]
	\centering
    \begin{overpic}[width=0.7\linewidth]{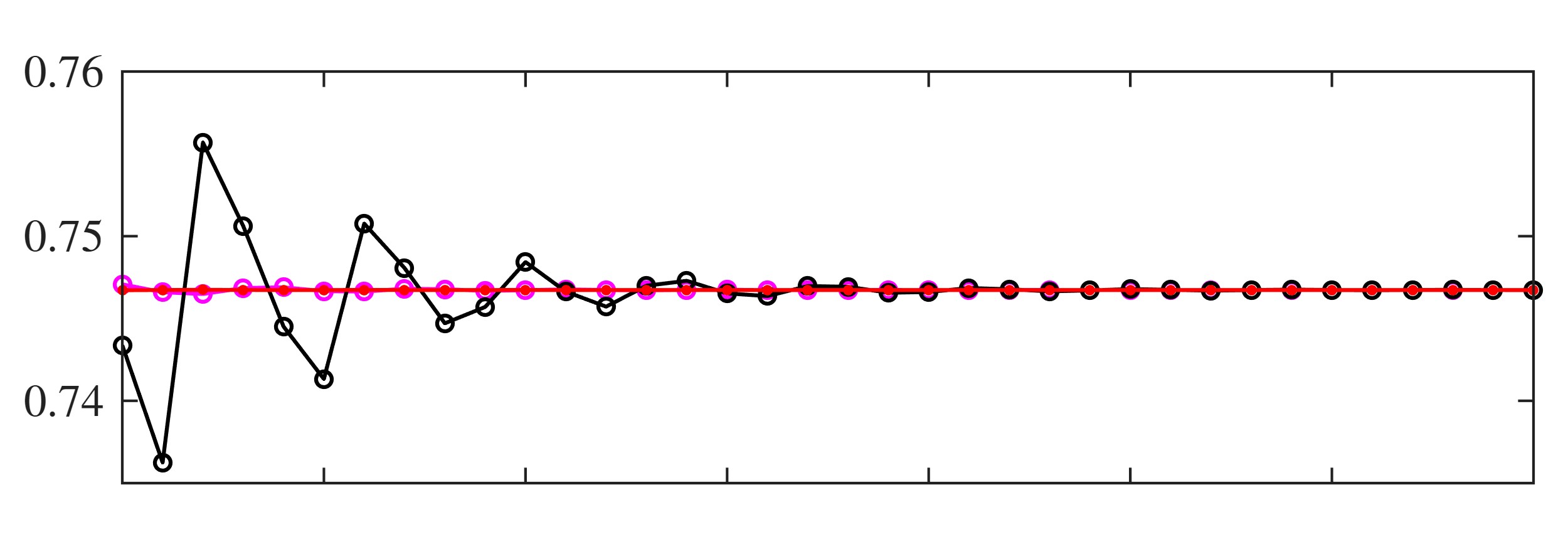}
		\put(-5, 30){(a)}
		\put(-6, 17){$ \omega_r $}
	\end{overpic}
    \begin{overpic}[width=0.7\linewidth]{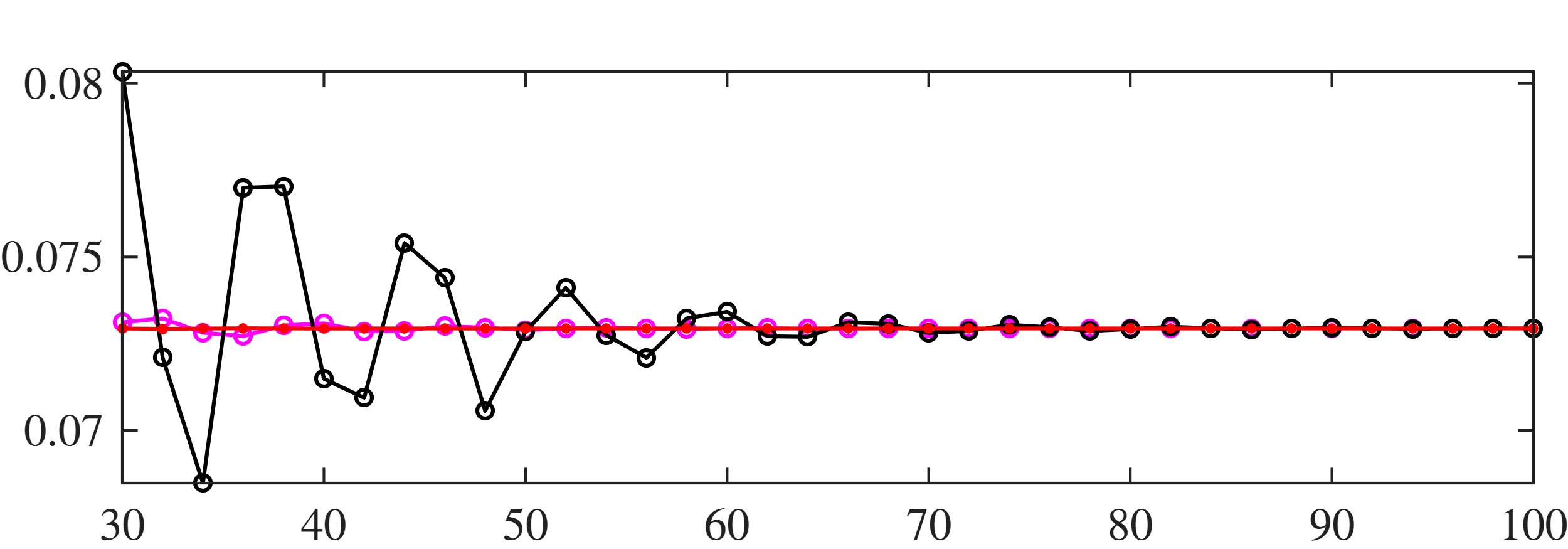}
		\put(-5, 30){(b)}
		\put(-6, 17){$ \omega_i $}
        \put(52, -1.5){$ x_\text{o} $}
	\end{overpic}
\caption{Same caption as figure~\ref{fig:spectrum-different-BCs-cylinder-40} for $ \R = 70 $ in the current plots. Different boundary conditions are denoted by different colors: Dirichlet (magenta), Neumann (black), and Robin-C (red).}
	\label{fig:spectrum-different-BCs-cylinder-70}
\end{figure}
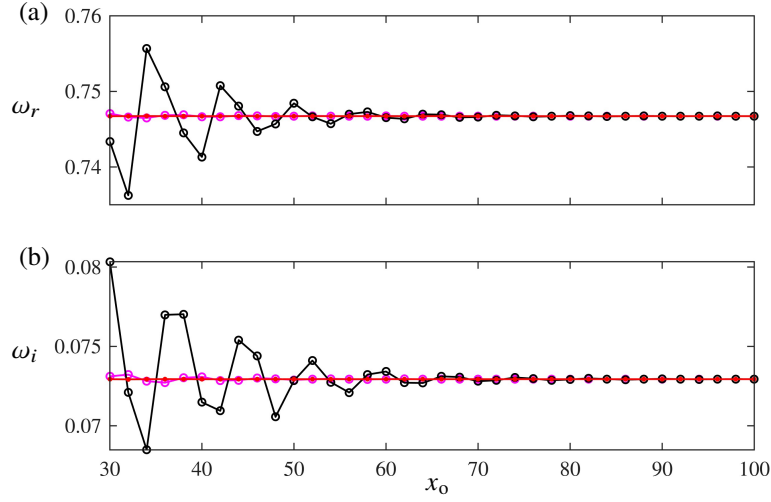

Figure~\ref{fig:spectrum-different-BCs-cylinder-70} shows the eigenvalue of the most unstable mode of the cylinder wake at $\R=70$, with colors representing different BCs consistent with figure~\ref{fig:spectrum-different-BCs-cylinder-40}. Only the Neumann (black), Dirichlet (magenta) and Robin-C  red) conditions are shown (the extrapolation method, stress-free, and Robin-R conditions are excluded for brevity, as explained in the preceding sections). The growth rate $\omega_i$ changes its sign from negative to positive as $\R$ increases from 40 to 70, validating that the flow becomes globally unstable at the higher Reynolds number, consistent with the results in \citet{barkley2006linear} and \citet{mittal2010stability}.

The eigenvalues calculated from Neumann and Dirichlet BC show a significant improvement in convergence compared to the lower $\R$ case in figure~\ref{fig:spectrum-different-BCs-cylinder-40}. The difference between the classical Neumann or Dirichlet and Robin BC shows a clear trend with $\xo$, past $ \xo=60 $ the eigenmode is approximately converged for all OBCs. The fact that the eigenmode of all OBCs becomes converged at large $\xo$ for $\R=70$ indicates that (at current $\R$) if the computational domain is large enough, the global stability analysis becomes less sensitive to the choice of OBCs, and therefore we do not need to worry about how to choose the boundary condition as long as the outlet is far enough from the cylinder, which is consistent with \citet{dong2014robust} and others.

However, this behavior holds only for the higher-$\R$ case considered here. At low $\R$, selecting an appropriate OBC remains essential regardless of the domain length, and simply extending the computational domain does not ensure spectral convergence. This is somewhat counterintuitive, as one might expect the higher-$\R$, globally unstable case to be more demanding in terms of accuracy, leading to the common assumption that the domain length should scale linearly with $\R$. In contrast, the current results indicate that it is the low-$\R$ regime in which convergence is more difficult to achieve and more sensitive to the choice of OBCs for the current global stability analysis. This apparent discrepancy can be explained by examining the spatial evolution of the global eigenmodes.

Figure \ref{fig:u-mode-cylinder-Re-70} presents the real part of the eigenfunction components $\hat{u}$, $\hat{v}$, and $\hat{p}$ for the most unstable mode at $\R = 70$, obtained using Neumann (a--c), Dirichlet (d--f), and Robin (g--i) outlet boundary conditions, with the outlet placed at $\xo = 52$. The velocity components are nearly indistinguishable across all three OBCs, indicating that any of them would be adequate for this particular combination of $\R$ and $\xo$. Consequently, the velocity modes alone do not allow a meaningful comparison of convergence among the different OBCs. In contrast, the pressure component exhibits significantly stronger oscillations than the velocity components---most notably for the Neumann condition in panel (c). The striped pattern in the Dirichlet case (f) further reveals that, although the eigenvalues appear to converge for $\xo \approx 40$, the pressure field remains poorly converged near the outlet. The Robin condition again shows no visible oscillations, as seen in panel (i), in stark contrast to panels (c) and (f).

The most important difference in the eigenfunctions between $\R$=40 and $\R$=70 is their spatial development (\citet{mittal2010stability}). At $\R$=40 all eigenmodes tend to grow monotonically along the streamwise direction, as can be seen in figure~\ref{fig:spatial-growth-rate-alphai-cylinder-Re-40}. At $\R$=70, the eigenmodes no longer grow monotonically with $x$, instead they have a local maximum at $ x \approx 10$, past which the eigenmodes start to decay spatially and eventually die off, as shown in figure~\ref{fig:u-mode-cylinder-Re-70}. The direct comparison of the spatial growth of the global modes for the two different $\R$ can be found in~\ref{app:spatial-comparison}.

The differing spatial evolution of the eigenmodes explains the Reynolds-number dependence of the OBC behavior discussed above. At low $\R$, the eigenmodes exhibit spatial growth, making both the Neumann and Dirichlet conditions incompatible with the actual physics at the outlet. Increasing the domain length does not alter this mismatch and therefore does not improve convergence (see also eq.~\eqref{eq:U-straight} and the accompanying discussion). At higher $\R$, however, the eigenmodes decay downstream, and their amplitude becomes negligible at sufficiently large $x$. As a result, both Neumann and Dirichlet conditions become increasingly consistent with the true downstream behavior, explaining why their performance improves with outlet location for high~$\R$.

From the comparison of eigenvalues and eigenfunctions at the two Reynolds numbers, two conclusions emerge. First, the Robin boundary condition consistently outperforms the classical Neumann and Dirichlet conditions for both stable and unstable cylinder wakes, yielding noticeably better convergence of both the spectrum and the eigenfunctions—particularly the pressure component. Second, the influence of domain length on OBC performance depends strongly on the Reynolds number: Whereas increasing the domain size is effective for unstable flows at high $\R$, it offers limited improvement for the stable, low $\R$ regime, at least for the present global stability analysis.

This can be explained by the relationship between spatial and temporal growth rates in the dispersion relation near the neutral point. At low $\R$, the negative temporal growth rate corresponds to the positive spatial growth rate, which cannot be captured by the Neumann and Dirichlet BC at the outlet; while at high $\R$, the positive temporal growth rate corresponds to the negative spatial growth rate, leading to spatial decay in the wake, and therefore all OBCs tend to work better when the outlet is far enough (the mode decays substantially). In other words, the $\R$ dependence of the OBC choice depends on the actual spatial and temporal behavior of the problem, and the same sign between the spatial and temporal growth rate indicate that the impact of outlet BC is more prominent when the flow itself is spatially growing (temporally stable), for which the long computational domain does not help to provide a more physical boundary condition, and therefore the more physical Robin OBC is necessary. 

\begin{landscape}	
	\begin{figure}
		\begin{overpic}[height=9.5cm]{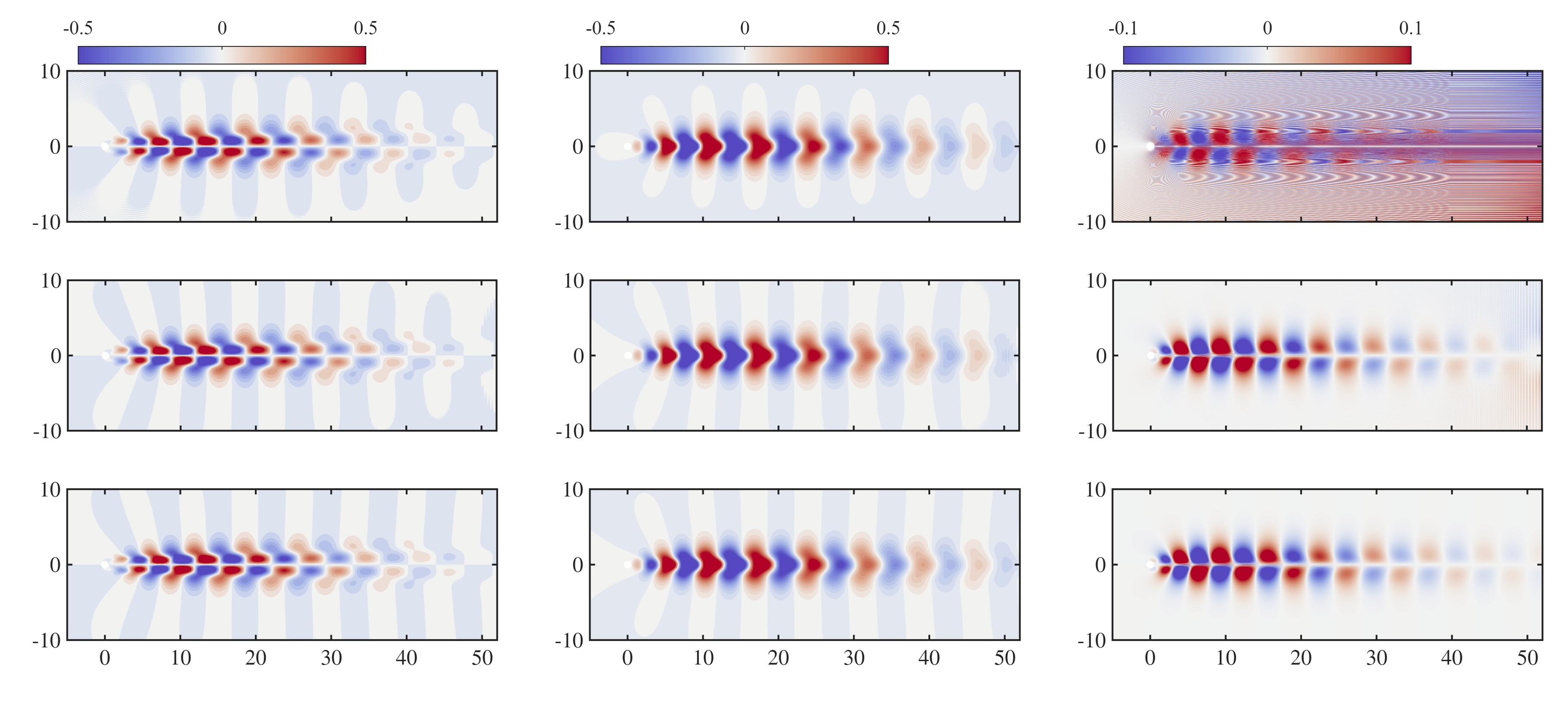}
			\put(0, 40){(a)}
			\put(33.3, 40){(b)}
			\put(67, 40){(c)}
			\put(0, 27){(d)}
			\put(33.3, 27){(e)}
			\put(67, 27){(f)}
			\put(0, 13.8){(g)}
			\put(33.3, 13.8){(h)}
			\put(67, 13.8){(i)}

			\put(24, 41.3){$\Re(\hat{u})$}
			\put(57.2, 41.3){$\Re(\hat{v})$}
			\put(90.8, 41.3){$\Re(\hat{p})$} 

			\put(1, 35.8){$y$}
			\put(1, 22.5){$y$}
			\put(1, 9.2){$y$}

			\put(19, 1){$x$}
			\put(52, 1){$x$}
			\put(85, 1){$x$}
		\end{overpic}
		\caption{Real part of the most unstable global eigenmode of the flow past a circular cylinder at $\R=70$: $ \hat{u}$ (first column), $ \hat{v}$ (second column), and $ \hat{p}$ (third column), for different outflow boundary conditions: Neumann (a,b,c), Dirichlet (d,e,f), and Robin (g,h,i). The outlet is located at $\xo = 52$. All eigenfunctions are normalized by the magnitude of the $\hat{u}$ mode at $(x,y)\approx (14, 1)$. The relative error, defined in eq.~\eqref{eq:gamma-p}, is $ \varepsilon_{\hat{p}}$ = 1.67 (c), 1 (f), and 0 (i).}
		\label{fig:u-mode-cylinder-Re-70}
	\end{figure}
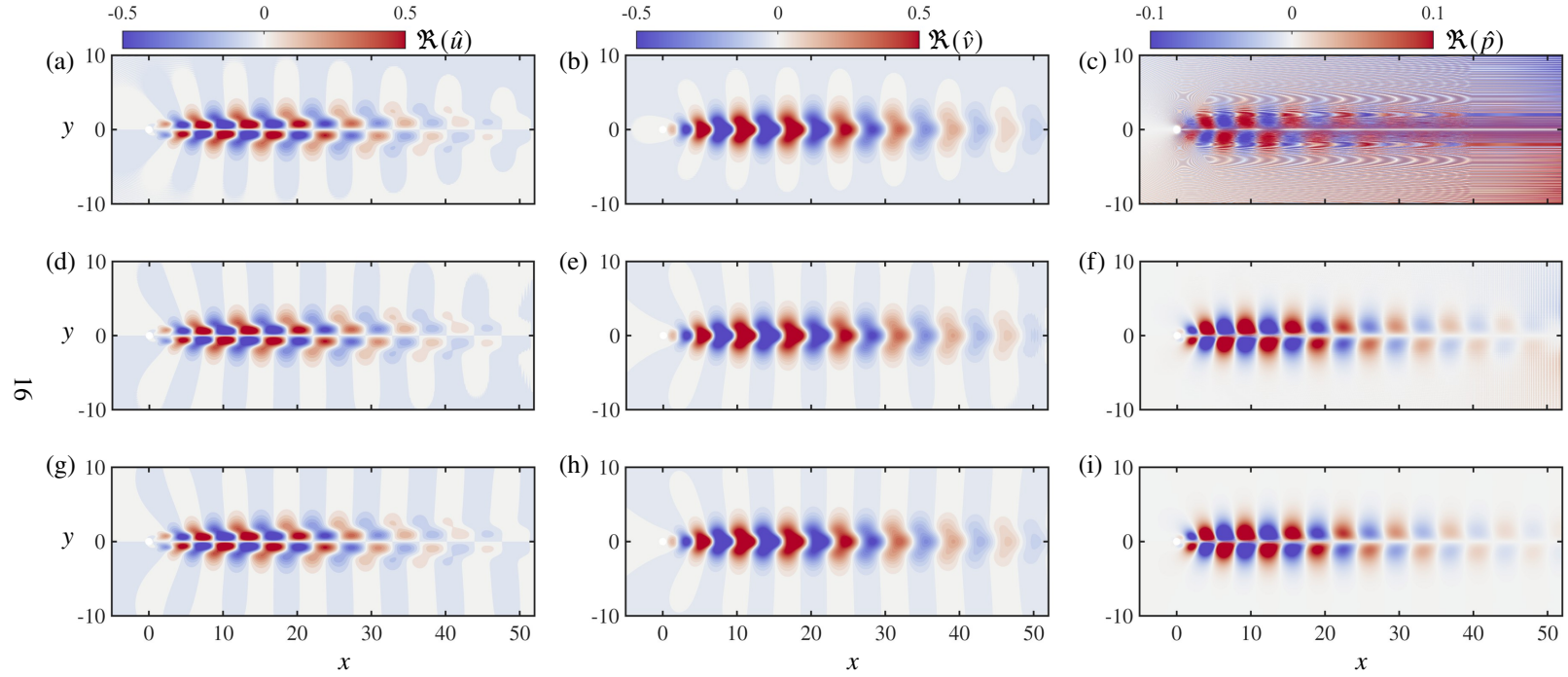	
\end{landscape}	 

\begin{figure}
	\centering
	\begin{overpic}[width=0.475\linewidth]{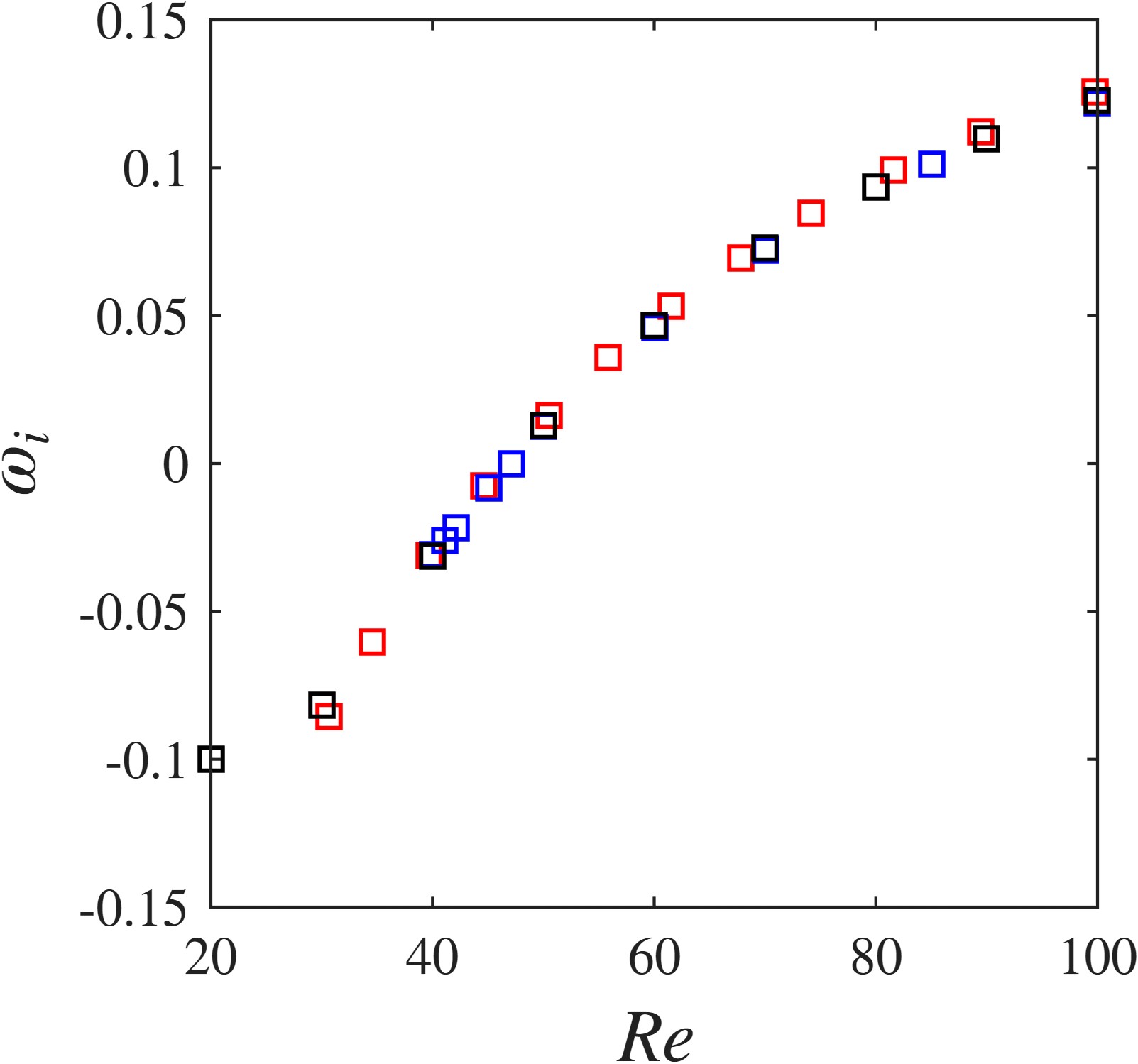}
		\put(-2, 90){(a)}
	\end{overpic}
	\hspace{2mm}
	\begin{overpic}[width=0.475\linewidth]{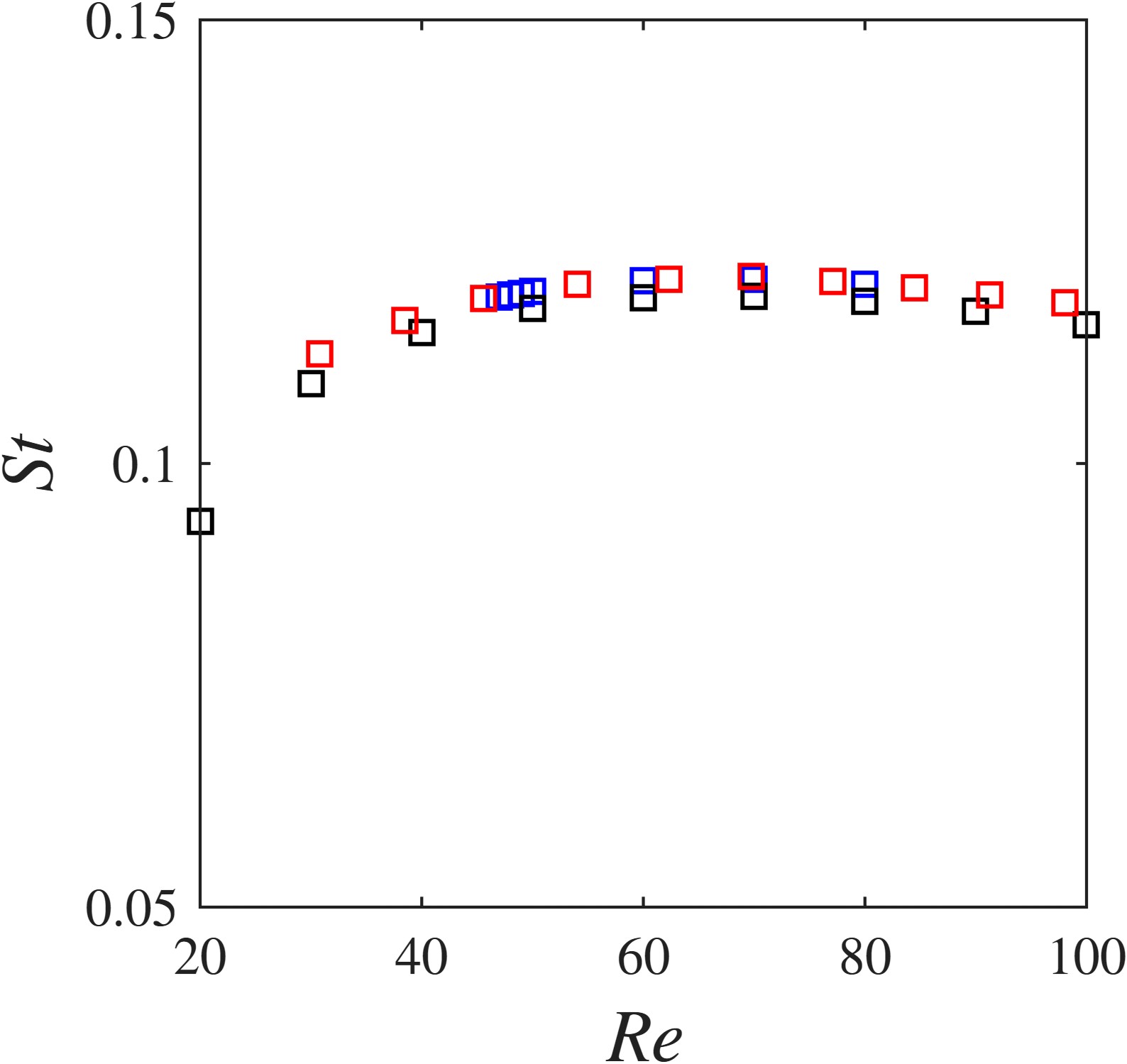}
		\put(-2, 90){(b)}
	\end{overpic}
	\caption{(a)~Growth rate, $\omega_i$, and (b)~Strouhal number, $St = \omega_r/2\pi$, variation with Reynolds number in the cylinder wake flow. Red: \citet{barkley2006linear}; blue: \citet{mittal2010stability}; black: present study.} 
	\label{fig:St-Re-effect-cylinder}
\end{figure}

Figure~\ref{fig:St-Re-effect-cylinder} shows the real and imaginary parts of the temporal spectrum with $\R$ ranging from 20 to 100, calculated from the complex Robin condition. The black squares, representing the current study, show a good match with the results in the literature for both globally stable $(\omega_i<0)$ and unstable $(\omega_i>0)$ cases.

\subsection{OBCs for Global Stability Analysis in Airfoil Wakes}	
\label{sec:airfoil}

In this section, we examine the influence of different OBCs on the global stability analysis of the less-studied airfoil wake, in order to assess the robustness of the Robin condition without prior knowledge of the flow’s stability.

The base flow of a NACA0015 airfoil at $\R = 200$ at an angle of attack of $18^\circ$ was studied as a representative case. The Reynolds number is defined as $\R = U_\infty c / \nu$, where $c$ denotes the airfoil chord length. These parameters were chosen to match the configuration studied in \citet{he2017linear}. A curvilinear C-grid was generated in Pointwise v18.1, with the wake extending $100c$ downstream and spanning $40c$ in the transverse direction, resulting in a computational mesh of $2257 \times 305$ points. The near-airfoil mesh is shown in figure~\ref{fig:grid-cylinder}(c). The global stability analysis was carried out using the same solver employed for the cylinder wake; additional details of the DNS and global stability methodology are provided in~\S\ref{sec:SU2}.

\begin{figure}
	\centering
	\begin{overpic}[width=0.85\linewidth]{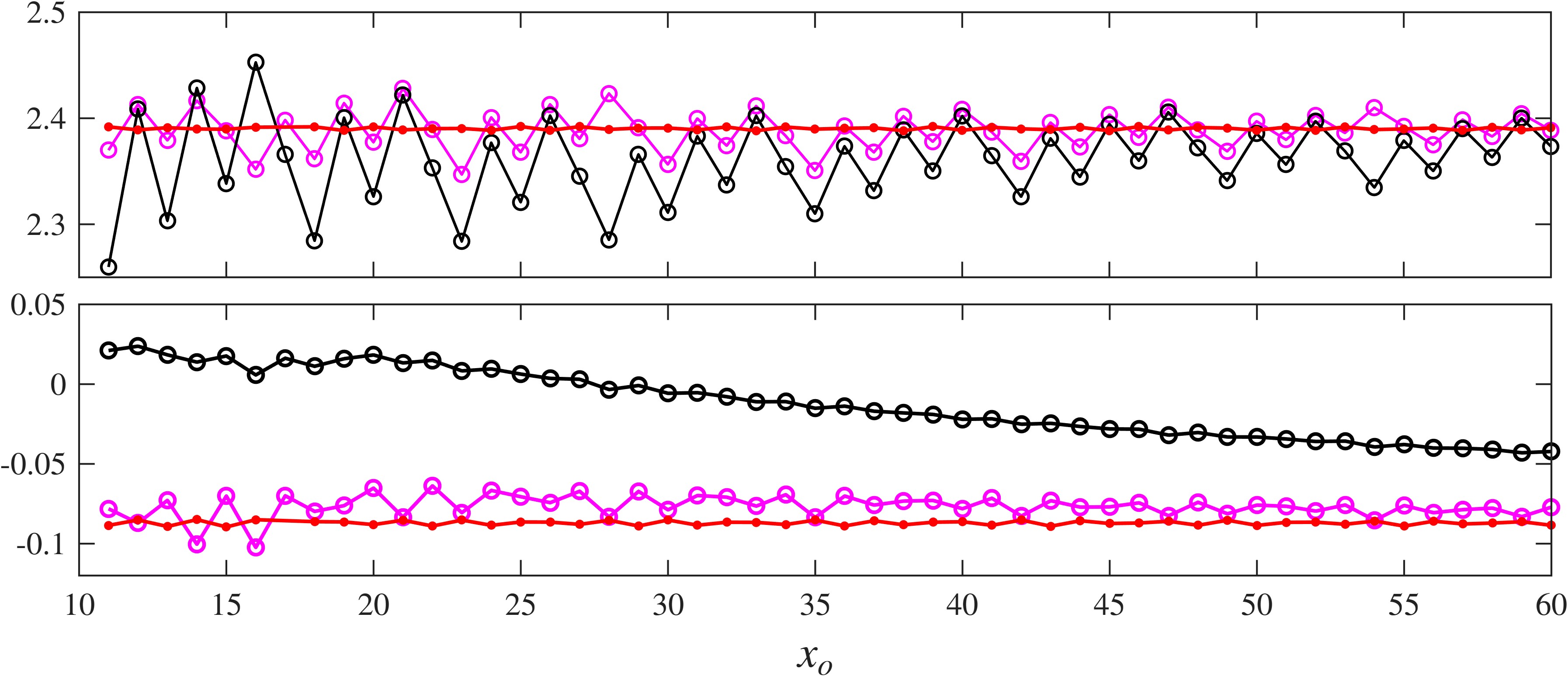}
		\put(-4, 41){(a)}
		\put(-4, 22){(b)}
		\put(-3.5, 14.1){$\omega_i$}
		\put(-3.5, 34){$\omega_r$}	
	\end{overpic}
\caption{Real~(a) and imaginary~(b) parts of the least stable eigenvalue of global stability analysis for NACA0015 airfoil with an angle of attack of $18^\circ$, at $\R = 200$, with the outlet location $ 10 < \xo < 60 $. Different OBCs are represented by different colors: Dirichlet (magenta), Neumann (black), and Robin-C (red).} 
	\label{fig:omega-different-BCs-airfoil-200}
\end{figure}
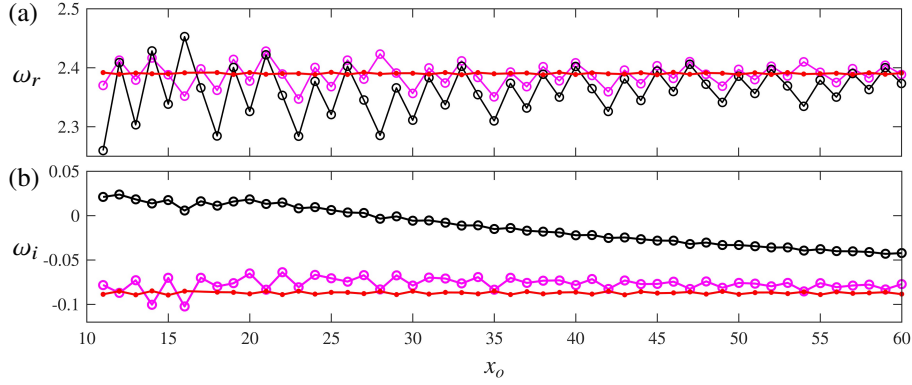

Figure~\ref{fig:omega-different-BCs-airfoil-200} presents the least stable eigenvalue of the airfoil wake for outlet positions $10 \leq \xo \leq 60$ under different OBCs. Similar to the cylinder case, the frequency $\omega_r$ obtained with the Neumann condition (black) exhibits pronounced spatial oscillations. Moreover, the growth rate $\omega_i$ in panel (b) shows that the Neumann condition yields an incorrect growth rate even for the largest domain considered ($\xo = 60$). As in the cylinder wake, the Dirichlet condition performs better than the Neumann condition; however, noticeable variations in both frequency and growth rate with respect to the outlet location remain.

The large variations in the spectrum with domain length are effectively mitigated by the Robin OBC, indicated by the red dashed line, even for a very small computational domain ($\xo \approx 10$). This demonstrates the robustness and efficiency of the Robin condition for flows past different bluff bodies, including cylinders and airfoils, allowing a significant reduction in computational domain size. A direct comparison of the global results as a function of domain length is further discussed in \S\ref{subsec:convergence-compare}.

Figure~\ref{fig:u-mode-airfoil-Re-200} shows the real part of the least stable eigenmode for different OBCs in the airfoil wake. The velocity components obtained with both the Neumann (first row) and Dirichlet (second row) conditions appear reasonable; however, closer inspection near the outlet reveals higher oscillations in $\hat{u}$ and $\hat{v}$ for the Dirichlet BC (d, e) compared to the Neumann BC (a, b). In contrast, the pressure component exhibits substantial distortion for the Neumann BC (c), whereas the Dirichlet BC (f) performs better for the pressure, despite showing stronger oscillations in the velocity components. Another notable difference is in the spatial behavior of the velocity modes within the computational domain: for the Neumann BC, their magnitude initially increases with $x$ and then decays slightly, whereas for the Dirichlet BC, the velocity magnitude increases monotonically with $x$.

These observations illustrate the complexity of OBC effects in global stability analysis, as different components of the eigenmode (velocity versus pressure) can respond differently to the same outlet condition. The oscillations in both velocity and, particularly, pressure components are effectively suppressed by the Robin OBC, as shown in the last row of figure~\ref{fig:u-mode-airfoil-Re-200}. The Robin BC captures the overall spatial trends similarly to the Dirichlet BC while eliminating high wavenumber oscillations near the outlet by enforcing a more physically consistent boundary condition.

\begin{landscape}
	\begin{figure}
		\begin{overpic}[height=9.8cm]{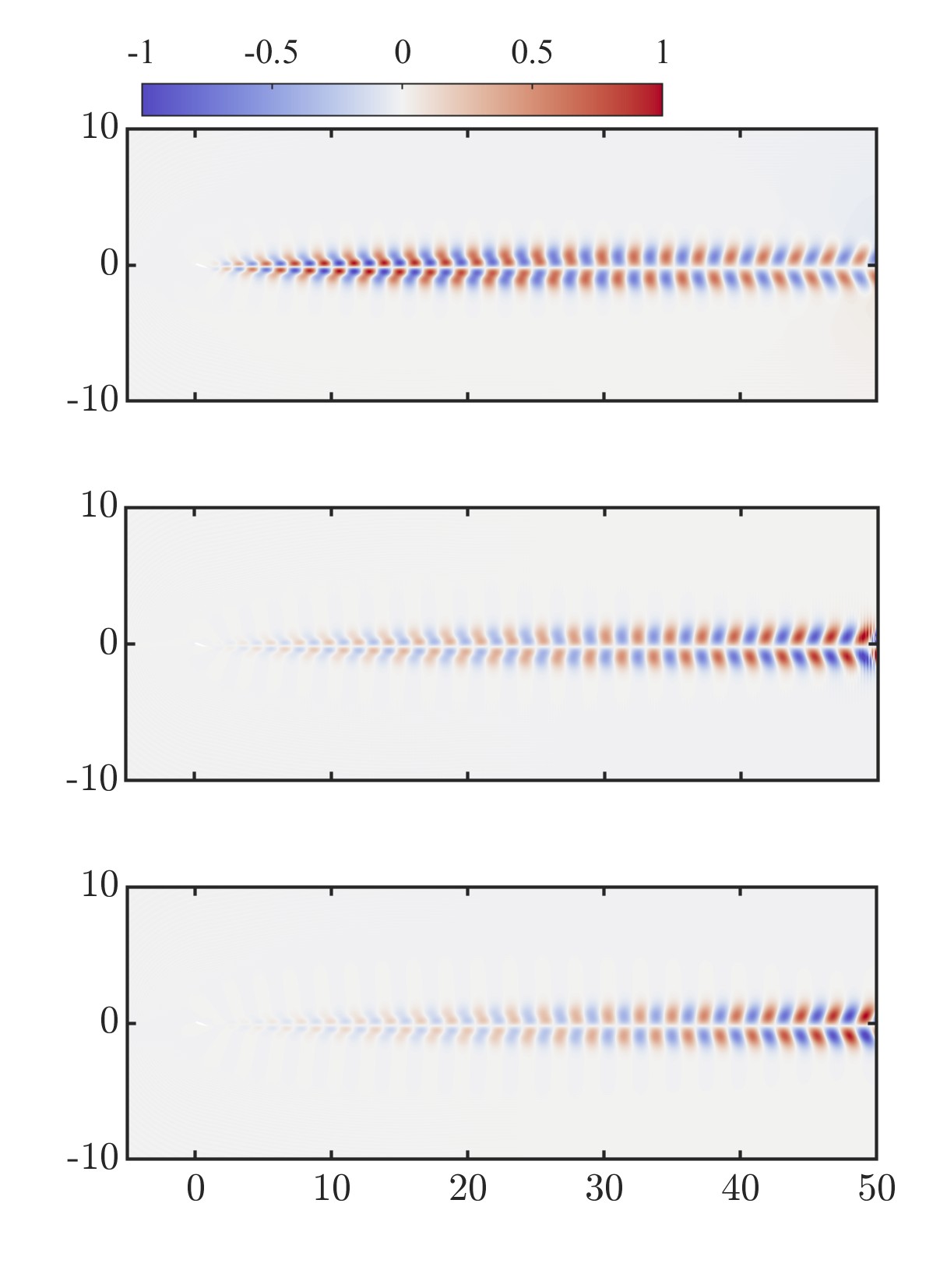}
			\put(0, 89){(a)}
			\put(0, 60){(d)}
			\put(0, 30){(g)}
			\put(2, 79){$y$}
			\put(2, 50){$y$}
			\put(2, 20){$y$}
			\put(41, 3){$x$}
			\put(53, 91){\scriptsize $\Re(\hat{u})$}
		\end{overpic}
		\begin{overpic}[height=9.8cm]{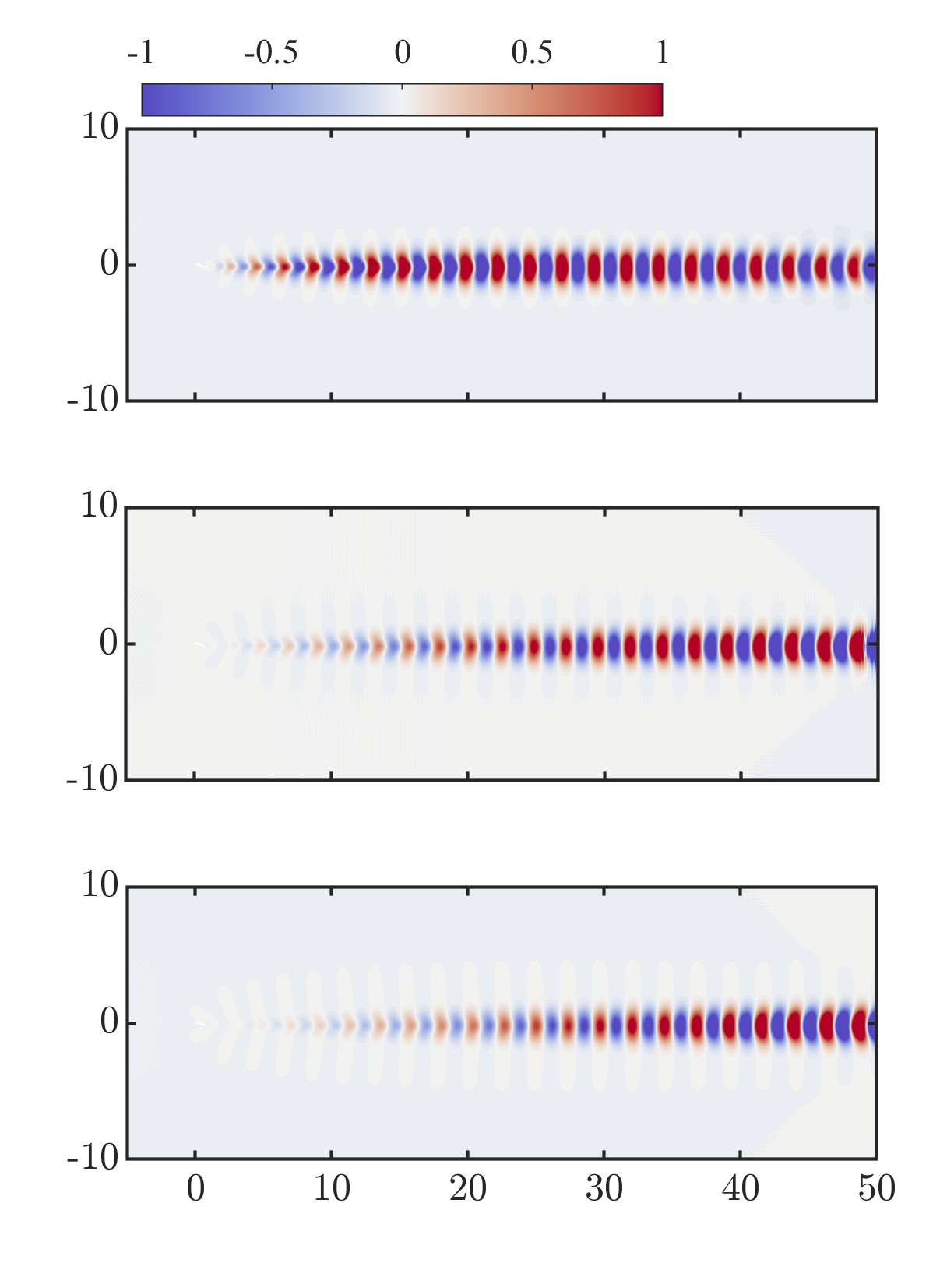}
			\put(0, 89){(b)}
			\put(0, 60){(e)}
			\put(0, 30){(h)}
			\put(41, 3){$x$}
			\put(53, 91){\scriptsize $\Re(\hat{v})$}
		\end{overpic}
		\begin{overpic}[height=9.8cm]{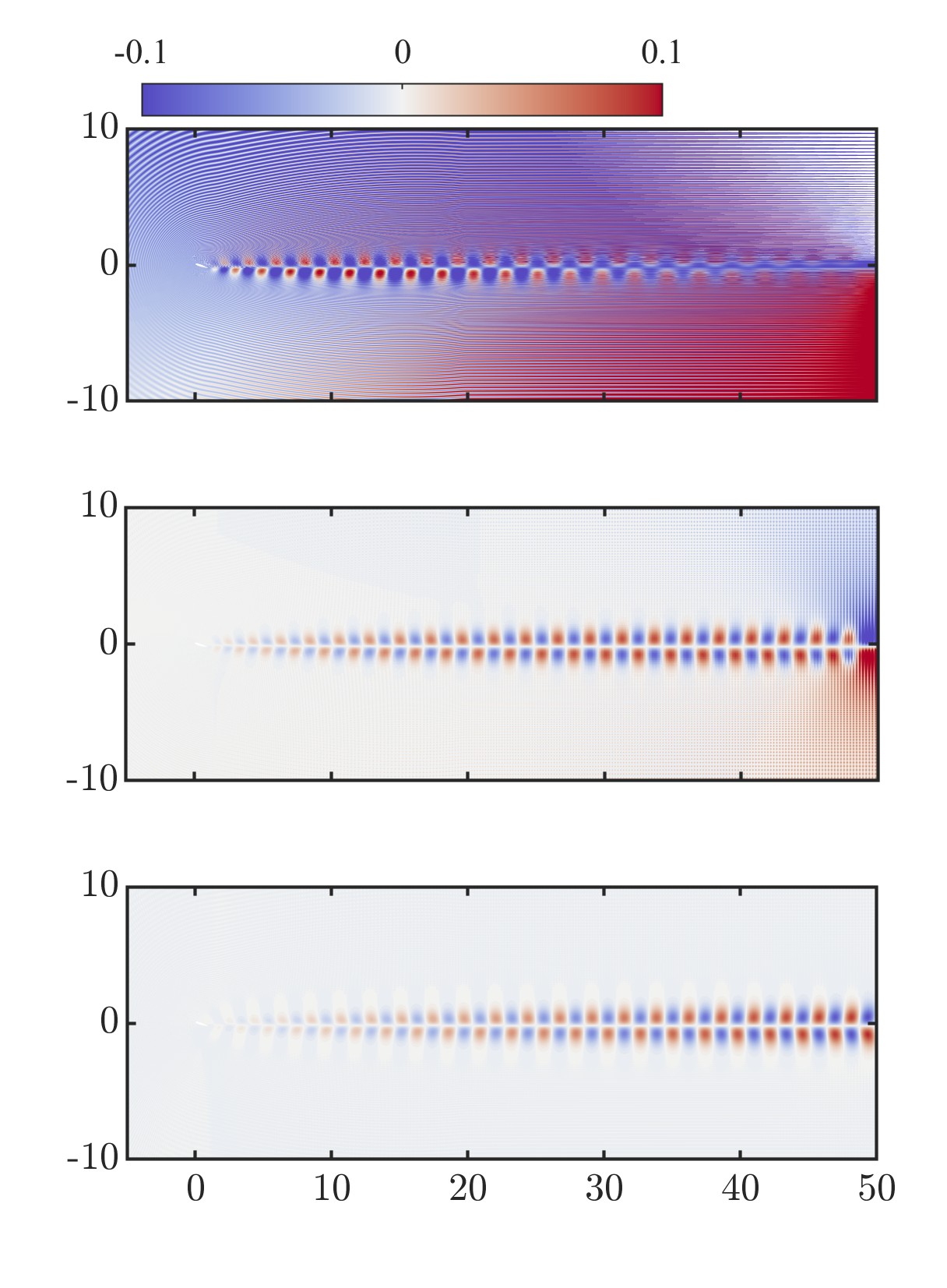}
			\put(0, 89){(c)}
			\put(0, 60){(f)}
			\put(0, 30){(i)}

			\put(41, 3){$x$}
			\put(53, 91){\scriptsize $\Re(\hat{p})$}
		\end{overpic}
\caption{Real part of the least stable eigenmode, $ \hat{u}$ (first column), $ \hat{v}$ (second column), and $ \hat{p}$ (third column), calculated from different outflow boundary conditions: Neumann (a,b,c), Dirichlet (d,e,f), and Robin (g,h,i) of NACA0015 airfoil with an angle of attack of $18^\circ$, at $ \R = 200 $. All eigenmodes are normalized by the magnitude of $\hat{u}$ at $(x, y) \approx (48, -1)$. The relative error, defined in eq.~\eqref{eq:gamma-p}, is $\varepsilon_{\hat{p}}$ = 1.04 (c), 1 (f), and 0 (i).}
		\label{fig:u-mode-airfoil-Re-200}
	\end{figure}
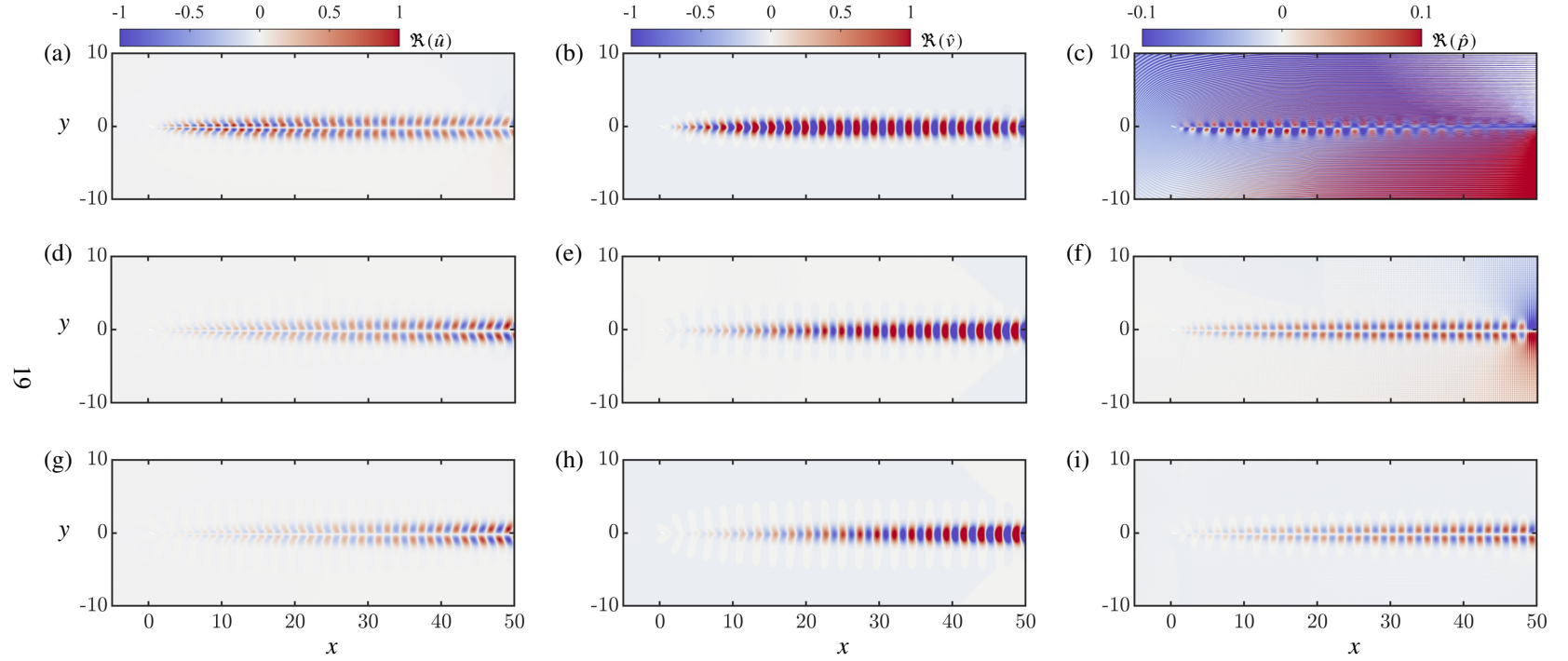	
\end{landscape}

\subsection{Stationary Modes in Airfoil Wake Flow}
\label{subsec:convergence-compare}

In this subsection, we examine the effectiveness of the Robin condition in capturing the stationary mode by performing the global stability analysis of the airfoil wake at different spanwise wavenumbers. 

\begin{figure}
\centering
	\begin{overpic}[width=0.95\linewidth]{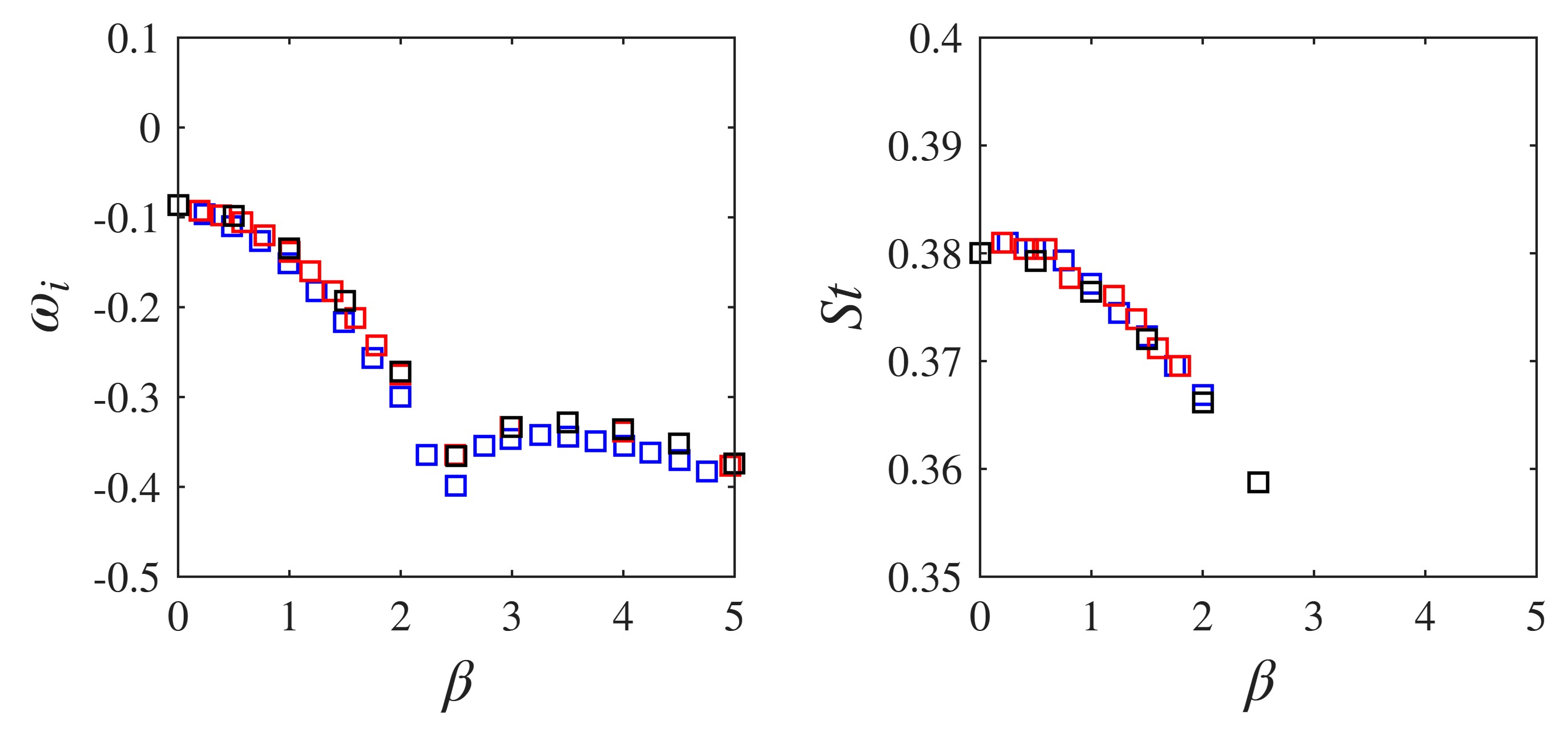}
		\put(0, 45){(a)}
		\put(52, 45){(b)}
	\end{overpic}
\caption{(a)~Growth rate, $\omega_i$, and (b)~Strouhal number, $St$, of the least stable modes of NACA0015 airfoil at $\R$=200 and angle of attack of $18^\circ$. Blue: \citet{he2017linear} with time stepping method (Nektar++) and domain $(\xo,y_{\max})=(50,15)$; red: \citet{he2017linear} with matrix-forming (FreeFEM++) method and domain $(\xo,y_{\max})=(40,32)$; black: present study with domain of only $(\xo,y_{\max})=(10,5)$.}
	\label{fig:beta-effect}
\end{figure}	

Figure~\ref{fig:beta-effect} shows the effect of the spanwise wavenumber, $\beta$, on the damping rate and frequency of the NACA0015 airfoil at $\R$=200 and angle of attack of 18$^\circ$ by implementing the Robin OBC in the global analysis. It validates that the spectrum in the present study matches well with \citet{he2017linear}, in which they applied both the time stepping (blue) and the matrix-forming methods (red), although the domain size in the current study is much smaller in both streamwise and transverse direction. The change of the least stable mode between the traveling and stationary mode is well captured in (a) at approximately $\beta = 2.5$. The effect of OBCs on the stationary mode was also consistent with the results in $\beta=0$ case shown in figure~\ref{fig:u-mode-airfoil-Re-200}, and the eigenmodes of different OBCs can be found in \ref{app:stationary-mode}.

Finally, we would like to compare the spectrum convergence as a function of computational domain size, and quantify how the current Robin OBC can significantly reduce the required domain size. Table~\ref{tab:results_comparison} lists the relative error, $\omega - \tilde{\omega}$, of both real and imaginary parts of the least stable mode, where $\tilde{\omega_r}$ and $\tilde{\omega}_i$ represent the real and imaginary parts at the largest domain size ($(\xo, y_{\max}) = (40, 10)$ presented herein). The spanwise wavenumber $\beta=1$ was chosen to make the direct comparison with the literature. For the same computational domains, the relative error in the current study is at least three orders of magnitude smaller than that of \citet{he2017linear} in which the Neumann OBC was implemented. For instance, even the smallest computational domain used in the present study, $(\xo, y_{\max}) = (10, 5)$, yields eigenvalues with a relative error of the order of $10^{-6}$ for both the damping rate and the frequency, as long as the more physical Robin condition is implemented properly at the outlet.

\begin{table}[ht]
\centering
\begin{tabular}{cc S[table-format=-2.3] S[table-format=-2.3] S[table-format=-2.3] S[table-format=-2.3]}
\toprule
 & & \multicolumn{2}{c}{\citet{he2017linear}} & \multicolumn{2}{c}{Present work (Robin OBC)} \\ 
\cmidrule(lr){3-4} \cmidrule(lr){5-6}
{$\xo$} & {$y_{\max}$} & {$|\omega_r - \tilde{\omega}_r| \times 10^{3}$} & {$|\omega_i - \tilde{\omega}_i| \times 10^{3}$} & {$|\omega_r - \tilde{\omega}_r| \times 10^{3}$} & {$|\omega_i - \tilde{\omega}_i| \times 10^{3}$} \\ 
\midrule
10 & 5	& {--} & {--} & {$4.8\times 10^{-3}$} & {$1.5\times 10^{-3}$} \\
15 & 5	& 6.7 & 9.1 & {$4.5\times 10^{-3}$} & {$1.2\times 10^{-3}$} \\
20 & 5	& 6.7 & 9.1 & {$4.5\times 10^{-3}$} & {$1.2\times 10^{-3}$} \\
30 & 5	& 4.3 & 9.8 & {$4.3\times 10^{-3}$} & {$1.2\times 10^{-3}$} \\
40 & 5	& 3.2 & 1.4 & {$4.5\times 10^{-3}$} & {$1.2\times 10^{-3}$} \\
30 & 10 & 4.4 & 9.8 & {$1.5 \times 10^{-8}$} & {$1.0 \times 10^{-9}$} \\
40 & 10 & {--} & {--} & {--} & {--} \\ 
\bottomrule
\end{tabular}
\caption{Effect of domain size on the global analysis for NACA0015 airfoil, $\R=220$, angle of attack of $18^\circ$ and spanwise wavenumber $\beta=1$, compared to Table 7 in \citet{he2017linear}. $\tilde{\omega}_i$, $\tilde{\omega}_r$ represent damping rate and frequency of the largest domain size reported in their study, $(\xo, y_{\max}) = (40, 10)$, listed in the last row.}
\label{tab:results_comparison}
\end{table}

\section{Conclusions}
\label{sec:conclusion}

The global stability analysis of the wake flow past bluff bodies was investigated for a wide range of computational domains using the matrix-forming method. The eigenvalues and eigenfunctions from the global analysis were found to be highly sensitive to the choice of the boundary condition at the outlet. Various outflow boundary conditions were compared in both cylinder wake and airfoil wake flows, and results show that the Robin boundary condition, which incorporates the local linear stability theory in the global solver, consistently yields a converged spectrum and a less distorted eigenfunction across different outlet truncations.

In the stable regime of the cylinder wake flow, the frequency calculated from the global stability analysis had high spatial oscillations for the classical Neumann, Dirichlet, extrapolation, stress-free, and sponge layer boundary conditions, and the damping rate showed even worse behavior, particularly for the Neumann and extrapolation conditions. Compared to the relatively smooth velocity modes, the pressure modes appeared to be much noisier and not well converged for both Neumann and Dirichlet conditions due to the non-physical condition enforced at the outlet. The Robin boundary condition was implemented by incorporating the local linear stability theory in the boundary condition of the global analysis, and a Taylor expansion was conducted near the neutral point to approximate the local streamwise wavenumber. We modified the previously reported Robin-R condition, which accounts only for the real part of the group velocity, and incorporated its full complex form (Robin-C) into the global stability solver. The latter enables us to capture more accurately the spatial growth of the global mode.

The effectiveness of the outflow boundary conditions depends on the spatial behavior of the eigenmodes at different Reynolds numbers. In the unstable regime of the cylinder wake, the spatially decaying nature of the modes makes the choice of OBC less affective with increasing computational domain, leading to a converged spectrum for all boundary conditions with sufficiently large domain size. In other words, the spectrum is easier to converge in the unstable regime, at higher $\R$. However, the lack of convergence of the spectrum at low $\R$ can not be overcome by simply extending the computational domain. The spatially amplifying nature of the modes at low $\R$ makes it essential to impose the physical Robin condition at the outlet. Therefore, global stability analysis is more sensitive to OBC when the flow is stable and the eigenmode is growing spatially {\textit{ad infinitum}}. In both globally stable and unstable cases examined in this study, the Robin condition consistently yields a converged spectrum with a much smaller domain size. 

The preference of Robin condition was further validated in the wake flow past the NACA0015 airfoil. The spanwise wavenumber trend of both damping rate and frequency in the present study matches well with both time stepping and matrix-forming results in \citet{he2017linear}. The convergence of the least stable mode was compared for the Robin condition and previously used Neumann conditions in the literature. The variation in the relative error of the spectrum was found to be about three orders of magnitude smaller for the same domain size compared to the results using the Neumann condition, showing that the computational cost can be significantly reduced when the more physical Robin outflow boundary condition is implemented.

The Robin outflow condition provides a physically consistent boundary condition for the global analysis of the bluff body wake flow. By incorporating the linear stability analysis for the local profile at the outlet, it has been shown that the Robin boundary condition can significantly reduce the computational domain required to obtain converged eigenmodes, particularly for low Reynolds number stable flows, for which classical boundary conditions cannot produce well-converged results simply by extending the domain size. With its clear physical mechanism and easy implementation in the matrix-forming framework, the Robin condition paves the way towards applications in more complex stability problems, such as Floquet analysis and in compressible flows.

\section*{Acknowledgment}
This research was supported by the ISRAEL SCIENCE FOUNDATION (grant No. 1522/21).

\appendix

\section{Metrics for Curvilinear Structured Grid}
\label{app:metrics}

The metrics in the eigenvalue decomposition for the current curvilinear C-grid were calculated using the finite difference method following \citet{anderson2002computational}. The transformation between the physical domain ($x, y$) and the computational domain ($\xi, \eta$) is given as

\be
\f{\p}{\p x}=\f{\p \xi}{\p x} \f{\p}{\p \xi} +\f{\p \eta}{\p x} \f{\p}{\p \eta}, \hspace{5mm} 
\f{\p}{\p y}=\f{\p \xi}{\p y} \f{\p}{\p \xi} +\f{\p \eta}{\p y} \f{\p}{\p \eta},
\ee
with the metrics given by

\be
\left[\begin{array}{ll}
\f{\p \xi}{\p x}	&	\f{\p \xi}{\p y} \\ [6pt]
\f{\p \eta}{\p x}	&	\f{\p \eta}{\p y}
\end{array}\right]=\f{1}{J}\left[\begin{array}{rr}
\f{\p y}{\p \eta} & -\f{\p x}{\p \eta} \\ [6pt]
-\f{\p y}{\p \xi} &	 \f{\p x}{\p \xi}
\end{array}\right],
\ee
where $J = x_\xi y_\eta - x_\eta y_\xi$ is the Jacobian of the transformation. Similarly, the second derivatives $\p ^2/\p x^2$, $\p ^2/\p y^2$ in the Laplacian operator have the following form \citep{anderson2002computational}

\bse
\be
\f{\p^2}{\p x^2} = 
\f{\p^2 \xi}{\p x^2} \f{\p}{\p \xi}
+ \f{\p^2 \eta}{\p x^2} \f{\p}{\p \eta}
+ \left(\f{\p \xi}{\p x}\right)^2 \f{\p^2}{\p \xi^2}
+ \left(\f{\p \eta}{\p x}\right)^2 \f{\p^2}{\p \eta^2}
+ 2 \f{\p \eta}{\p x} \f{\p \xi}{\p x} \f{\p^2}{\p \eta \p \xi},
\ee
\be
\f{\p^2}{\p y^2} = 
\f{\p^2 \xi}{\p y^2} \f{\p}{\p \xi}
+ \f{\p^2 \eta}{\p y^2} \f{\p}{\p \eta}
+ \left(\f{\p \xi}{\p y}\right)^2 \f{\p^2}{\p \xi^2}
+ \left(\f{\p \eta}{\p y}\right)^2 \f{\p^2}{\p \eta^2}
+ 2 \f{\p \eta}{\p y} \f{\p \xi}{\p y} \f{\p^2}{\p \eta \p \xi}.
\ee
\ese

\section{Base Flow Profiles in the Wake}
\label{app:wake-profile}

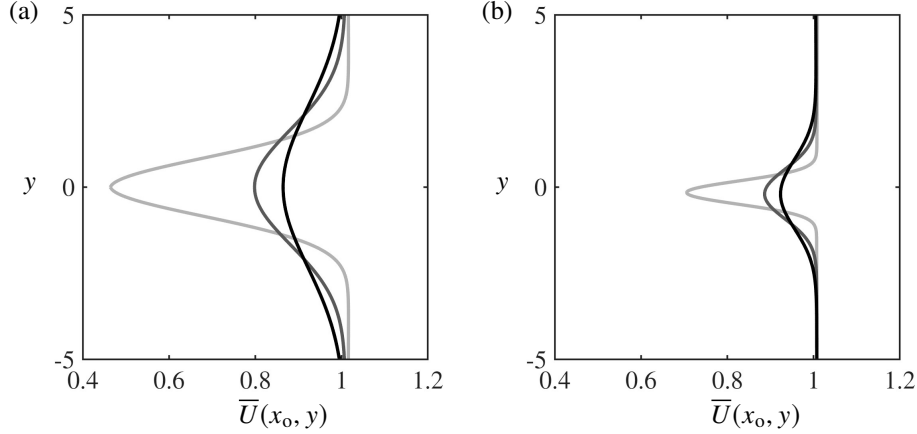
\begin{figure}[ht]
	\centering
	\begin{overpic}[width=0.45\linewidth]{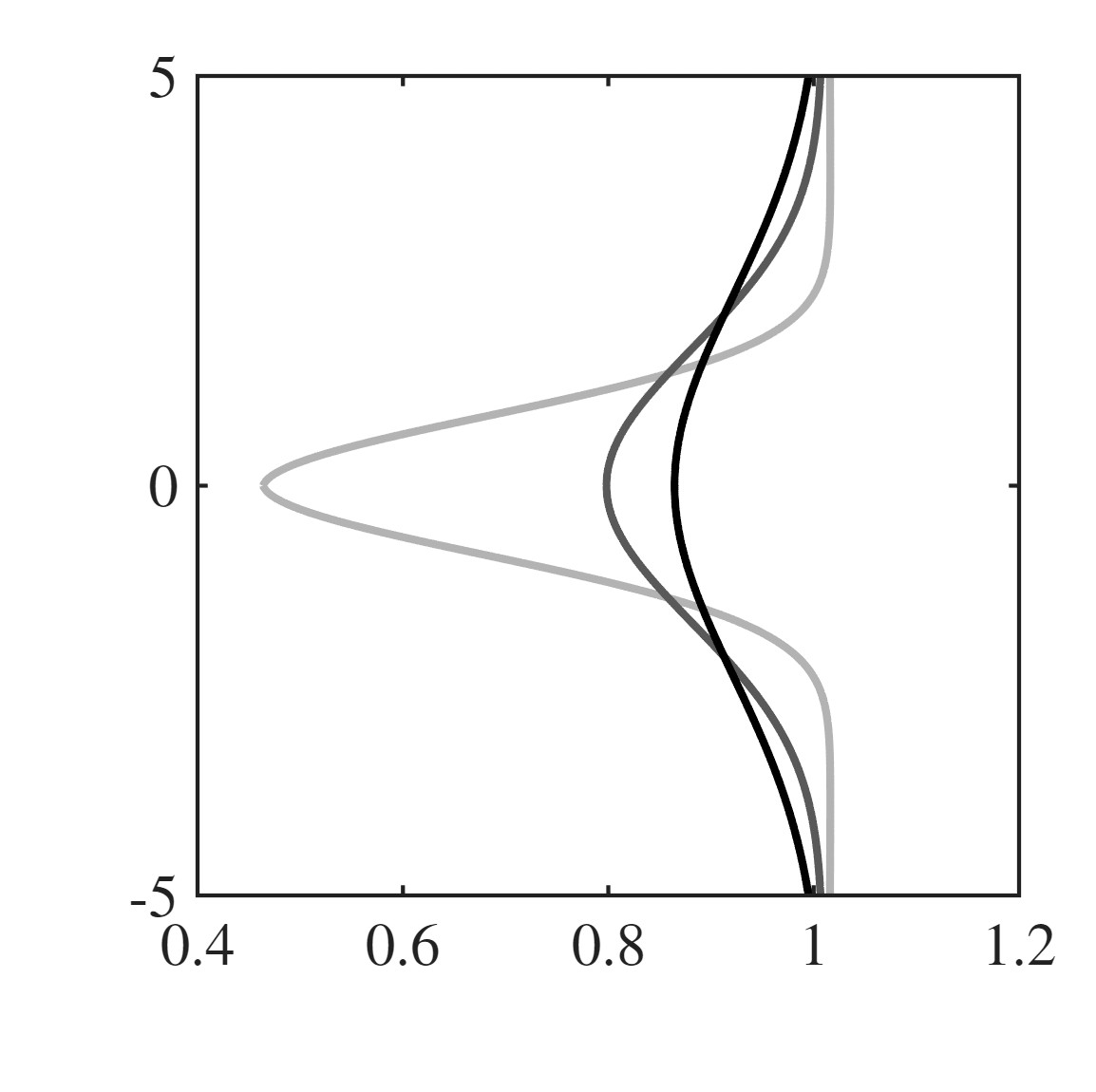}
		\put(2, 92){(a)}
        \put(5, 55){$y$}
        \put(52, 4){$\overline{U}(\xo,y)$}
	\end{overpic}
    \begin{overpic}[width=0.45\linewidth]{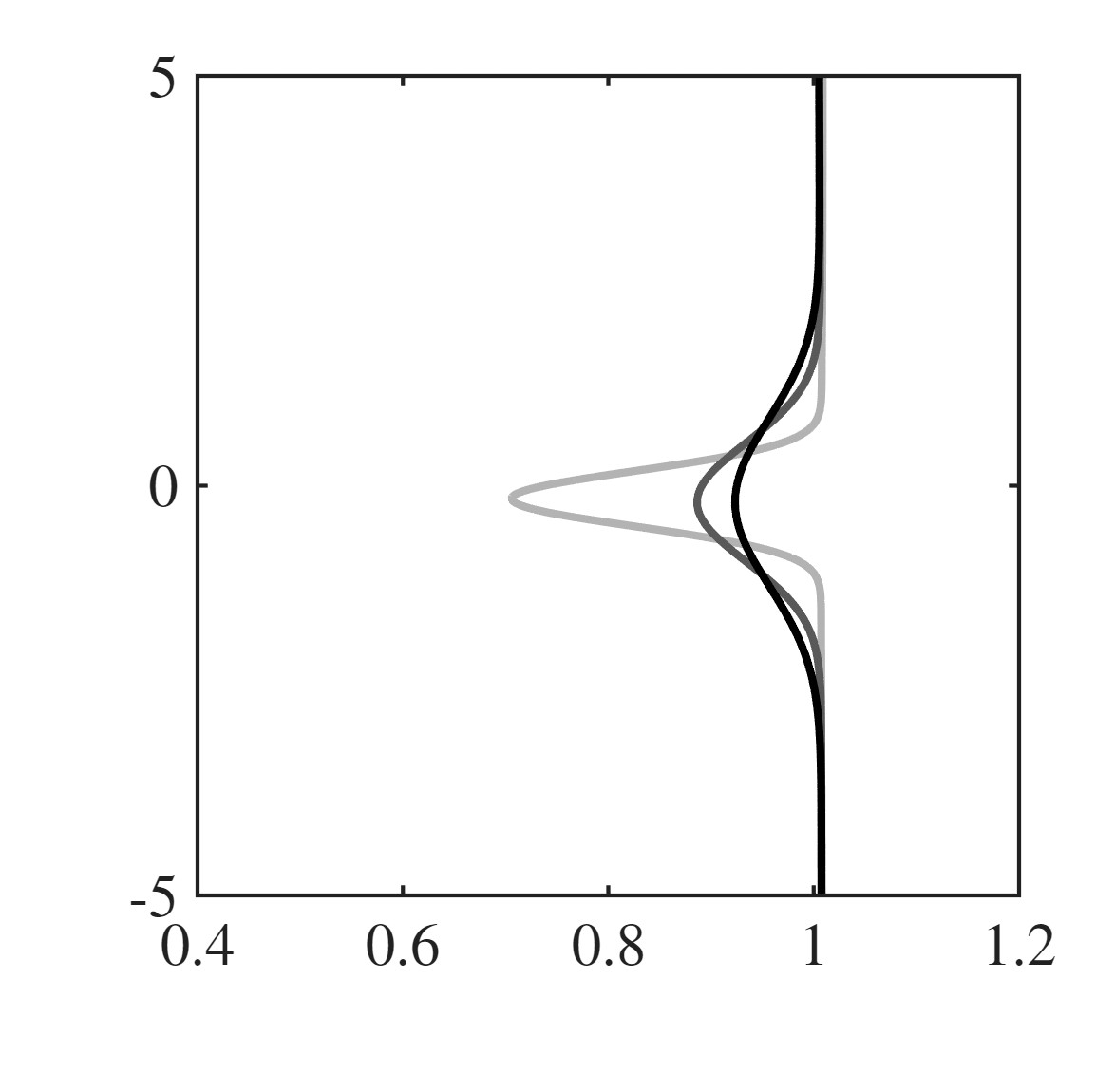}
		\put(2, 92){(b)}
        \put(5, 55){$y$}
        \put(52, 4){$\overline{U}(\xo,y)$}
	\end{overpic}
	\caption{Base flow streamwise velocity profile, $\overline{U}$, in the wake of (a)~Cylinder at $\R=40$, and (b)~NACA0015 airfoil at $\R=200$ and an angle of attack of $18^\circ$. The gray level of the lines represent profiles at $\xo=$ 10, 50, and 100, with darker lines representing larger $\xo$.}
	\label{fig:wake-U-profile}
\end{figure}

The base flow streamwise velocity profiles for the stable cylinder and airfoil cases are shown in Figure~\ref{fig:wake-U-profile} for several outlet locations $\xo$ = 10, 50, 100. As the wake develops downstream, the velocity deficit reduces and its thickness grows.

\section{Effect of Sponge Layers}
\label{app:convect-sponge}

The effect of a sponge layer at the outlet on the global stability analysis is analyzed for the case of the cylinder flow at $\R = 40$ as an example. A typical quadratic profile was chosen for the sponge,
\be
    \sigma(x) = \sigma_0 \left( \frac{x-x_\text{s}}{\xo - x_\text{s}} \right)^2  ,
\ee
where $x_\text{s}$ represents the starting location of the sponge, the sponge length $\lambda_\text{s} = \xo - x_\text{s}$, and $\sigma_0$ is the damping coefficient. We keep fixed $\lambda_\text{s} = 10$ for the sponge and vary the damping coefficient with a wide range $\sigma_0\in[0.05,1.5]$ following \citet{mani2012analysis}. 

\begin{figure}[ht]
	\centering
	\begin{overpic}[width=0.7\linewidth]{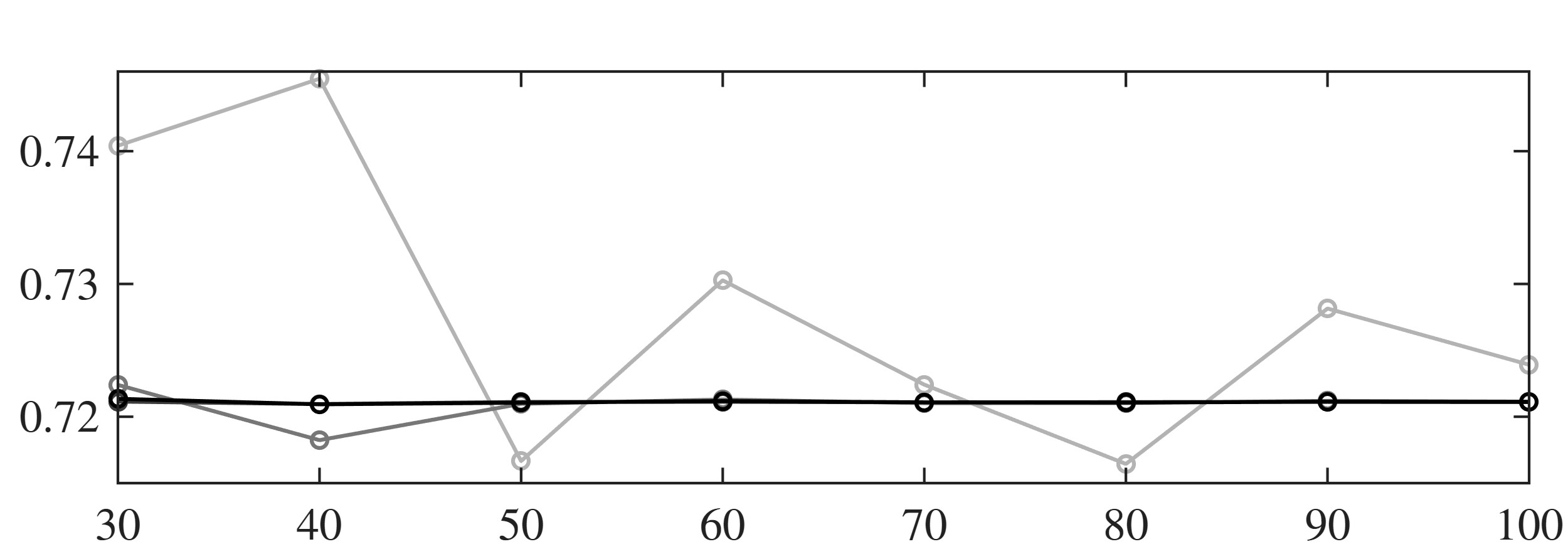}
		\put(-6, 29){(a)}
        \put(-5, 17){$\omega_r$}
	\end{overpic}
    \begin{overpic}[width=0.7\linewidth]{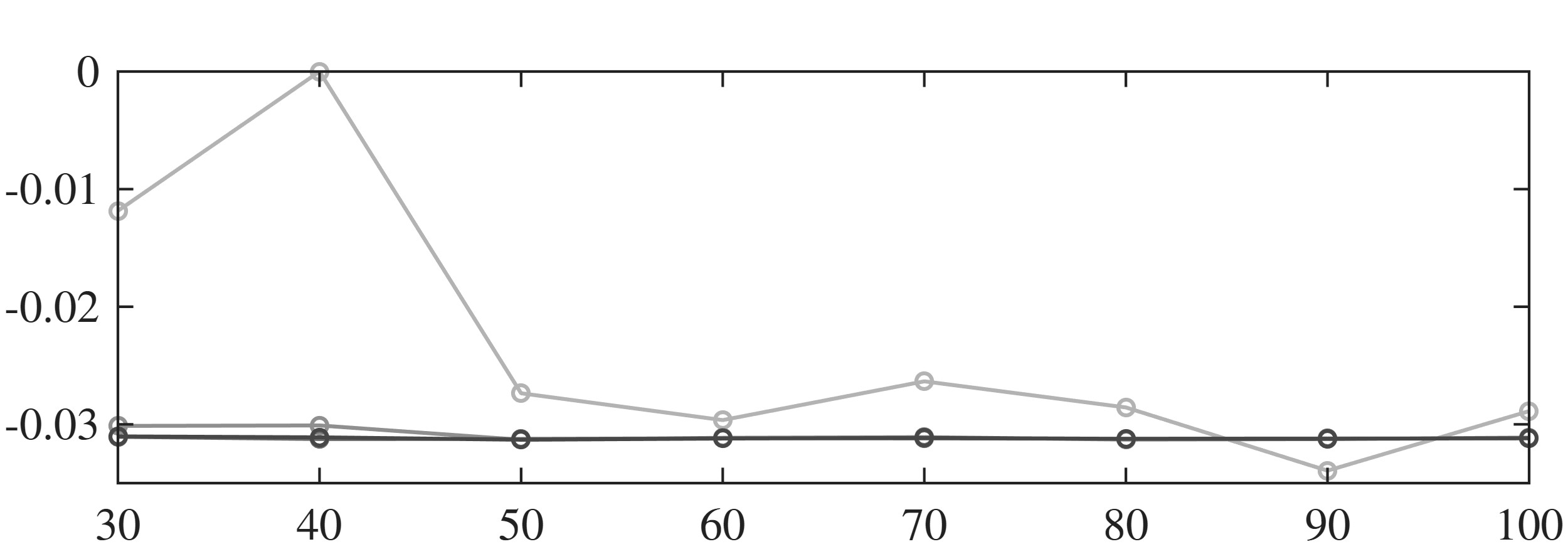}
		\put(-6, 29){(b)}
        \put(-5, 17){$\omega_i$}
        \put(52, -2){$x_\text{o}$}
	\end{overpic}
    \vspace{1mm}
\caption{Real~(a) and imaginary~(b) parts of $\omega$ for different damping coefficients of the sponge, with $\sigma_0 = $ [0.05, 0.2, 0.4, 1.5] corresponding to color levels from light to dark.}
	\label{fig:sponge-spectrum}
\end{figure}

\begin{figure}[ht]
	\centering
	\begin{overpic}[width=0.45\linewidth]{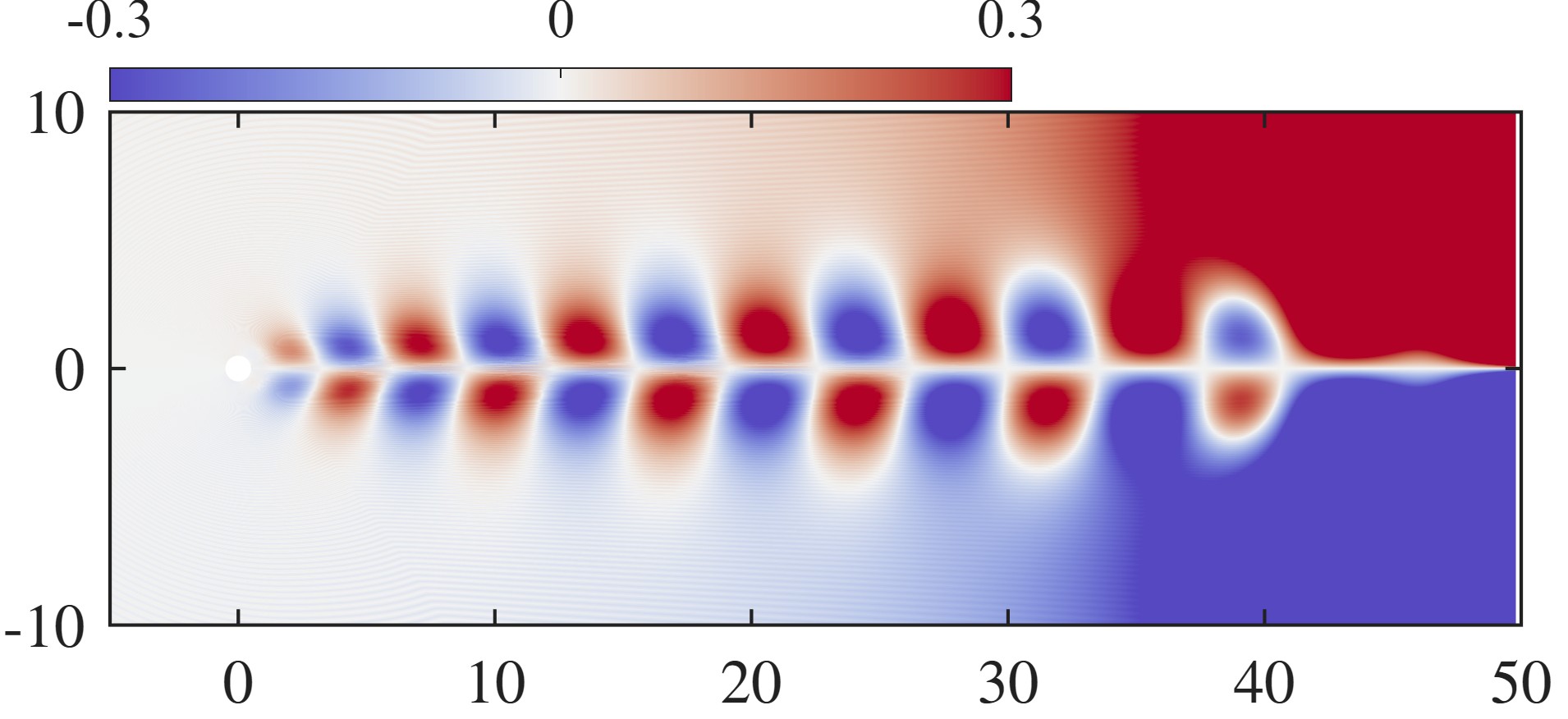}
		\put(-6, 38){\scriptsize(a)}
        \put(-5, 21){$y$}
        \put(68, 40){\scriptsize$\Re (\hat{p})$}
	\end{overpic}
    \hspace{2mm}
    \begin{overpic}[width=0.45\linewidth]{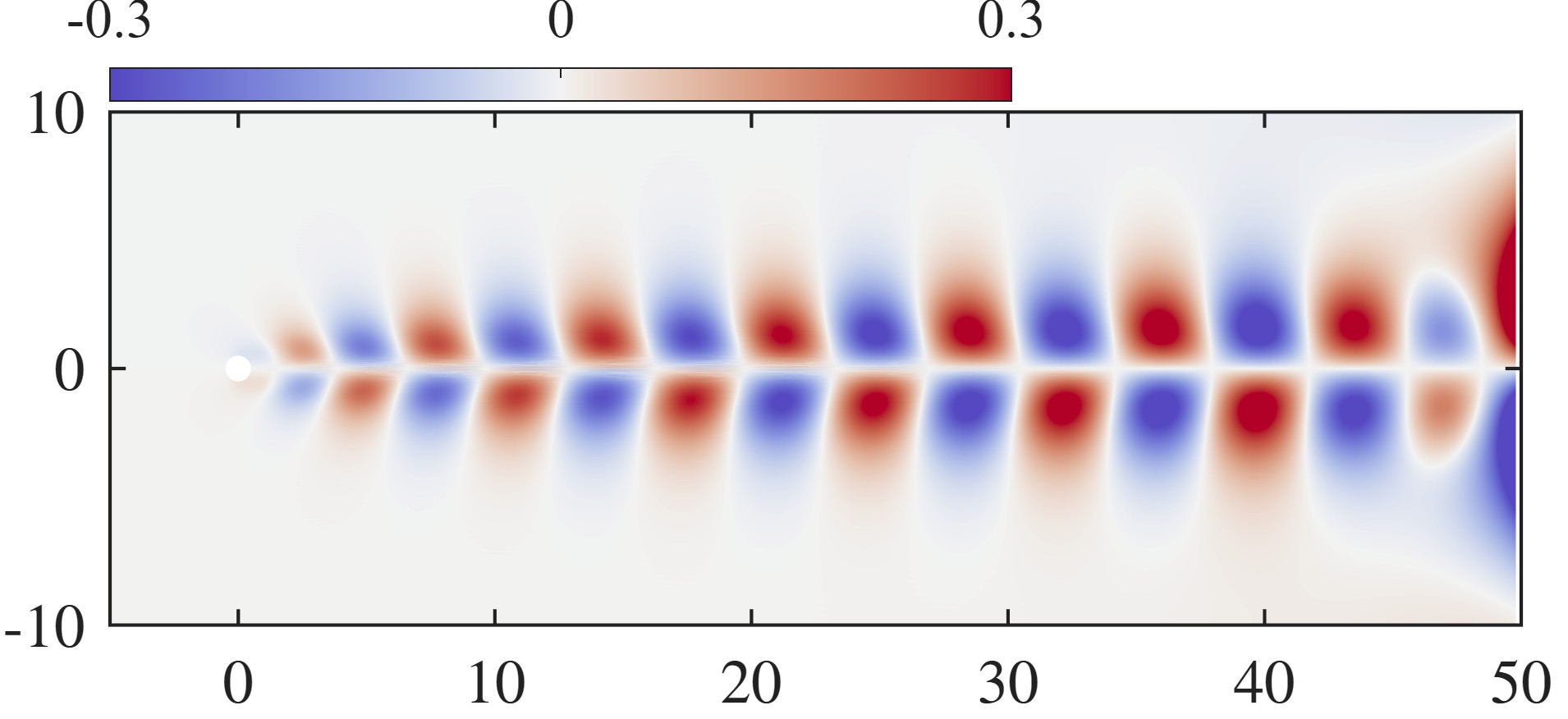}
		\put(-6, 38){\scriptsize(b)}
        \put(68, 40){\scriptsize$\Re (\hat{p})$}
	\end{overpic}

    \begin{overpic}[width=0.45\linewidth]{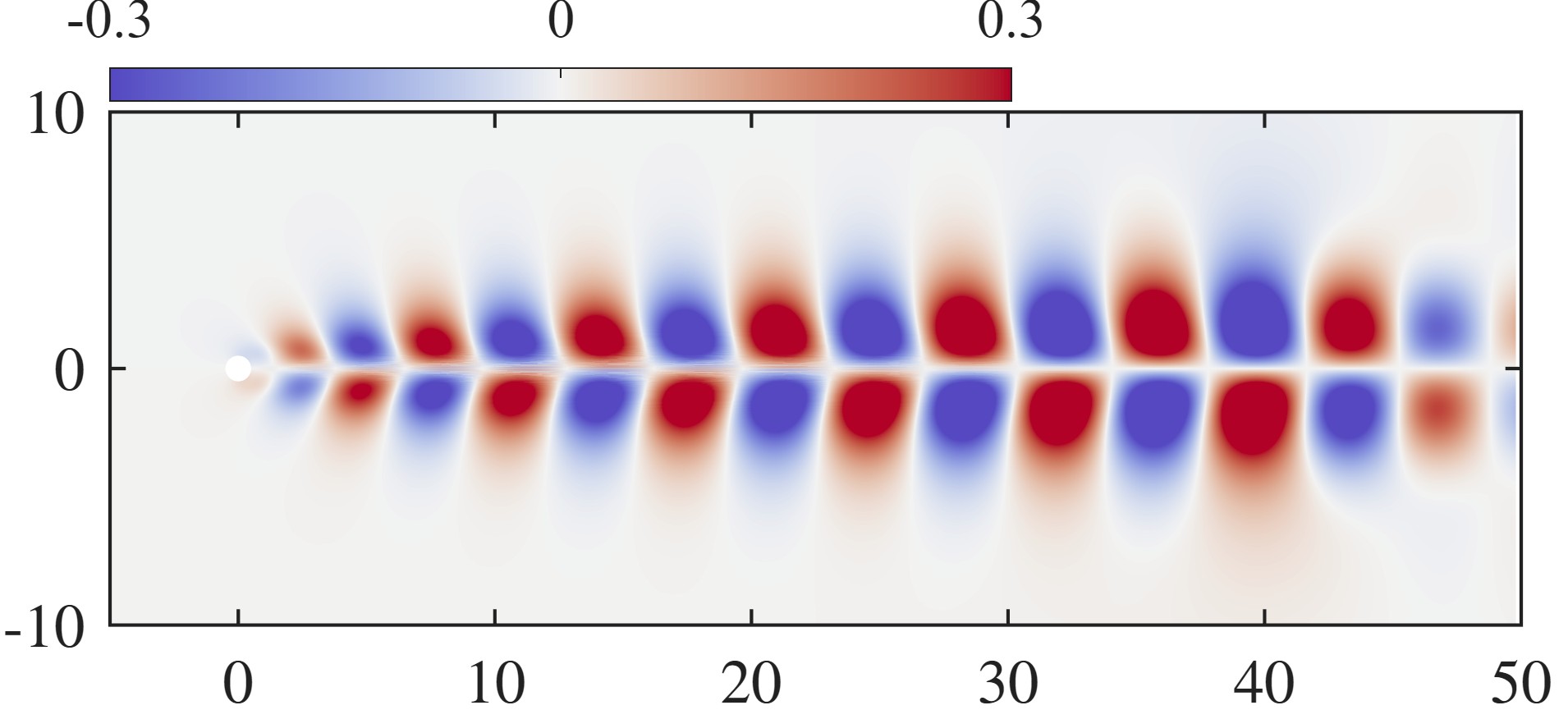}
		\put(-6, 38){\scriptsize(c)}
        \put(52, -3){$x$}
        \put(-5, 21){$y$}
        \put(68, 40){\scriptsize$\Re (\hat{p})$}
	\end{overpic}
\hspace{2mm}
    \begin{overpic}[width=0.45\linewidth]{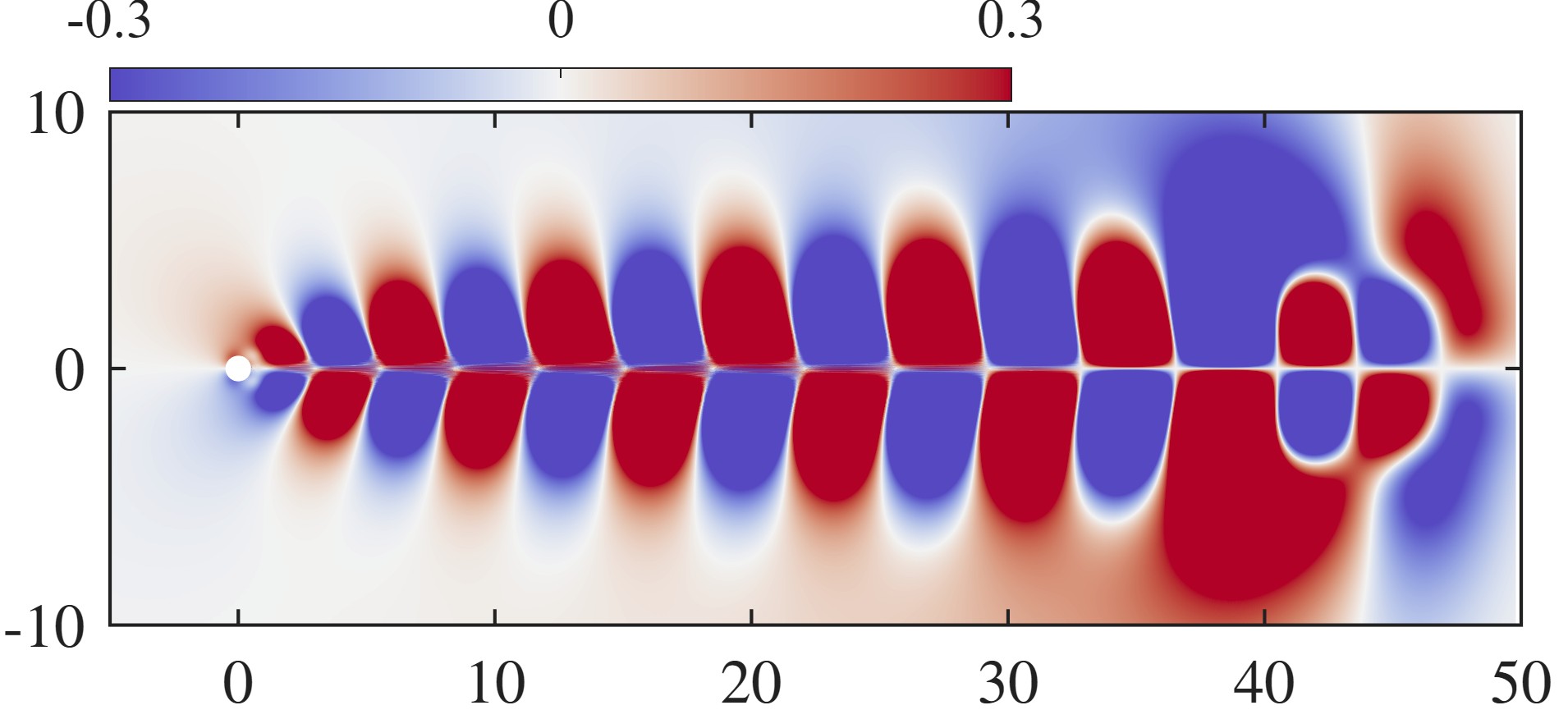}
		\put(-6, 38){\scriptsize(d)}
        \put(52, -3){$x$}
        \put(68, 40){\scriptsize$\Re (\hat{p})$}
	\end{overpic}
    
    \vspace{1mm}
\caption{Real part of pressure component of the least stable eigenmode, $\Re (\hat{p})$, for different damping coefficients of the sponge, shown in figure~\ref{fig:sponge-spectrum}, $\sigma_0 = $ [0.05, 0.2, 0.4, 1.5] for (a-d), respectively.}
	\label{fig:sponge-eigenmodes}
\end{figure}
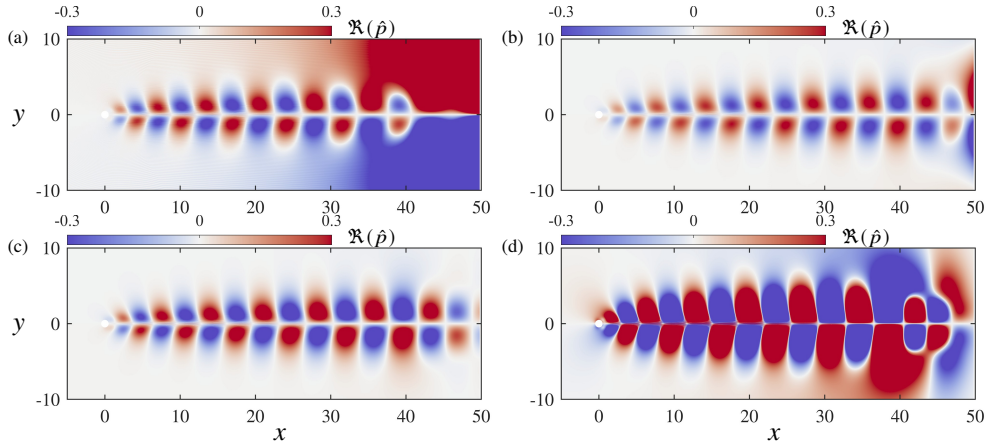

Figure~\ref{fig:sponge-spectrum} shows the variation of the least stable eigenvalue with $\xo$ for four different values of $\sigma_0$. It can be seen that the eigenmode convergence is highly dependent on the choice of damping coefficient. Similarly, figure~\ref{fig:sponge-eigenmodes} shows the pressure component of the least stable eigenmode corresponding to the same damping coefficient in figure~\ref{fig:sponge-spectrum}, and it is clear that the pressure component is highly dependent on the parameter choice. Therefore, without knowing \emph{a priori} the `optimal' damping coefficient, it would be difficult to determine whether the spectrum and eigenfunction are converged or not. Here we only consider the choice of $\sigma_0$, while in reality, the sponge length and actual damping profile would also highly affect the results, as described extensively in \citet{mani2012analysis}. This makes the implementation of the sponge layer outflow boundary condition unfavorable for general use in the current global stability analysis, as both the spectrum and eigenmodes can be highly dependent on the choices of the sponge layer parameters.

\clearpage

\section{Local Spatial Growth rate of the Global Mode at Different Reynolds numbers}
\label{app:spatial-comparison}

\begin{figure}[ht]
	\centering
	\begin{overpic}[width=0.8\linewidth]{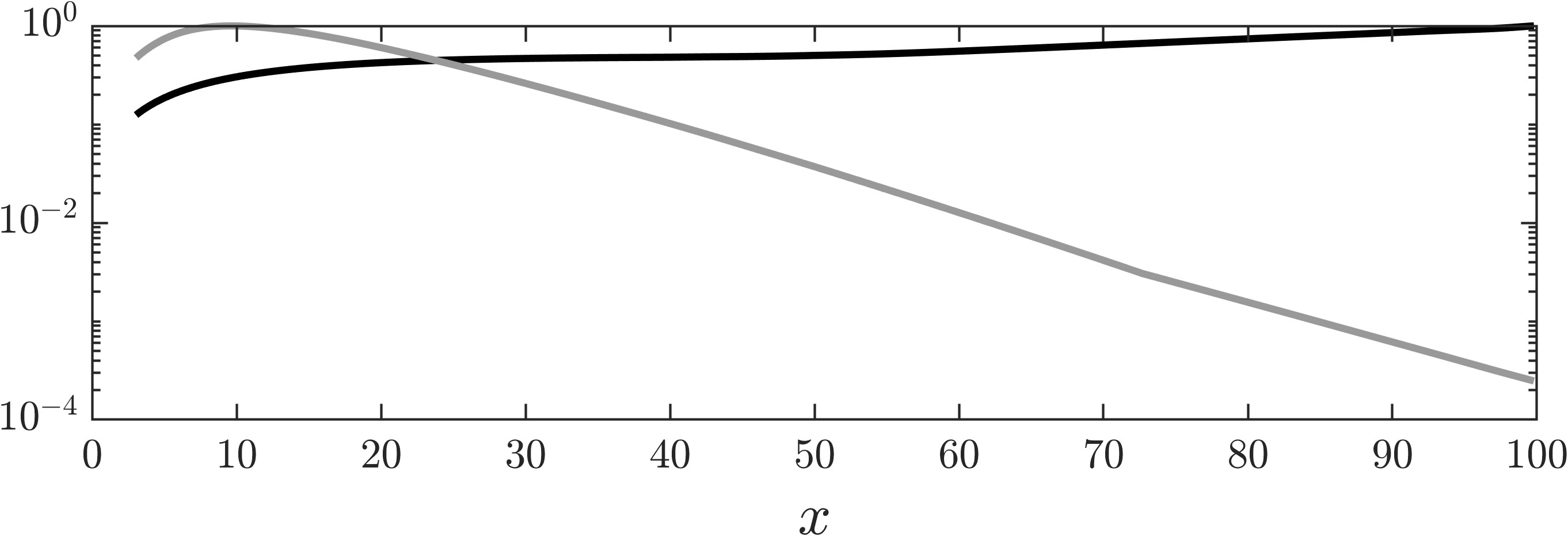}
		\put(-4, 15){\rotatebox{90}{$\text{max}_y|\hat{u}| $}}
	\end{overpic}
	\caption{Spatial development of the maximum magnitude of the streamwise velocity component, $\hat{u}$, (across all transverse points) at each streamwise location $x$ for $\R$ = 40 (black), and 70 (gray) of the cylinder wake flow.}
	\label{fig:spatial-growth-Re-40-70-u-mode}
\end{figure}

Figure~\ref{fig:spatial-growth-Re-40-70-u-mode} illustrates the comparison of the spatial global mode development for the two $\R$ of the cylinder wake flow. It is clear that at low $\R$ (black), the eigenmode tends to increase monotonically with $x$, while there is a local maximum for the higher $\R$ case (gray). The magnitude of the velocity component is normalized by the global maximum for each $\R$. This comparison implies that the choice of OBC affects more of the lower $\R$ case, because the mode tends to grow monotonically; while at higher $\R$, the mode itself decays very rapidly along $x$, and therefore it becomes less important which OBC we impose as long as the outlet location is large enough.

\section{Eigenmodes Comparison of Stationary Modes}
\label{app:stationary-mode}

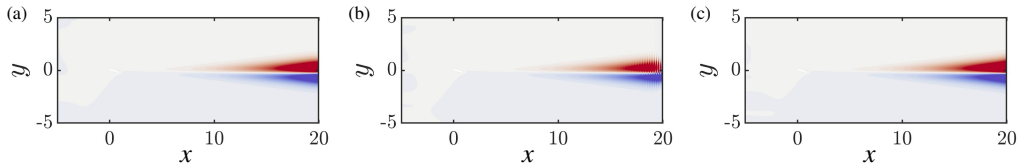
\begin{figure}[h]
\centering
	\begin{overpic}[width=\linewidth]{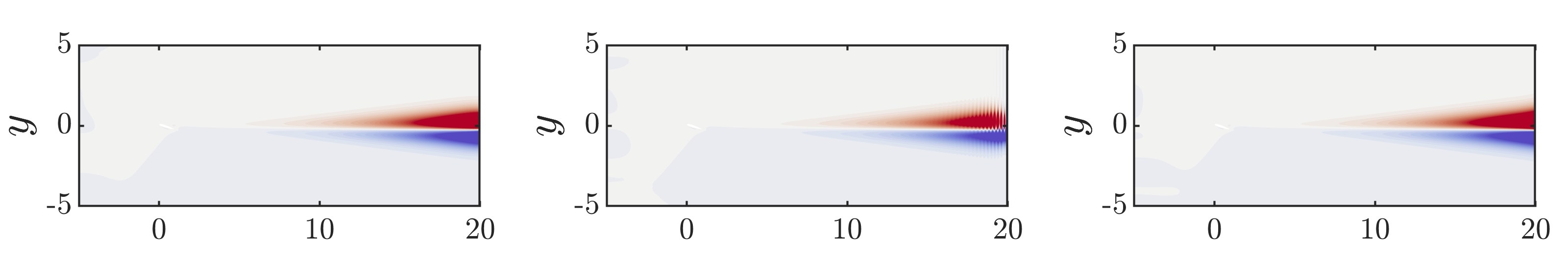}
		\put(0, 14){\scriptsize(a)}
		\put(33.5, 14){\scriptsize(b)}
		\put(67, 14){\scriptsize(c)}

		\put(17, 0){$x$}
		\put(50.5, 0){$x$}
		\put(84, 0){$x$}
	\end{overpic}
\caption{Real part of the spanwise velocity component, $\hat{w}$, of the least stable stationary global eigenmode at wavenumber $\beta = 3$ for the NACA0015 airfoil at $\R = 200$ and an angle of attack of $18^\circ$. Results are shown for (a)~Neumann, (b)~Dirichlet, and (c)~Robin outflow boundary condition. The outlet is located at $\xo=20$. Contour levels are normalized to range from $-$1 to 1.}
	\label{fig:stationary-mode}
\end{figure}

The eigenmode of the stationary mode for $\beta = 3$ is shown in figure~\ref{fig:stationary-mode} for the (a)~Neumann, (b)~Dirichlet and (c)~Robin outflow boundary condition. Only the spanwise velocity component, $\hat{w}$, is shown since other components ($\hat{u}$, $\hat{v}$, $\hat{p}$) are negligible compared to it. Interestingly, the $\hat{w}$ mode has the highest oscillation in the vicinity of the outlet for the Dirichlet condition in~(b), while it is much smoother throughout the computational domain for the Neumann~(a) and Robin~(c) condition, indicating that the Neumann condition performs better than the Dirichlet condition, which is opposite to the results at $\beta=0$ shown in figure~\ref{fig:u-mode-airfoil-Re-200}. Despite the distinct behaviors associated with the Neumann and Dirichlet boundary conditions, the Robin condition consistently produces smooth eigenmodes, thereby confirming its reliable and consistent application as an outflow boundary condition for the current global stability analysis.

\bibliographystyle{elsarticle-num-names}
\bibliography{bib}

\end{document}